\documentclass[twocolumn,prx,superscriptaddress]{revtex4}
\pdfoutput=1
\usepackage{amsmath,breqn,mathtools,amsfonts} 
\usepackage{hyperref}
\allowdisplaybreaks
\usepackage{graphicx}
\usepackage{tikz}
\usepackage{xcolor}
\usepackage{bm}
\usepackage{rotating}
\usepackage[export]{adjustbox}

\DeclarePairedDelimiterX{\ExpArg}[1]{[}{]}{#1}

\def\dbar{{\mathchar'26\mkern-12mu d}}

\makeatletter
\let\cat@comma@active\@empty
\makeatother

\begin{document}

\title{Signatures of irreversibility in microscopic models of flocking}

\author{Federica Ferretti}
\altaffiliation{Current address: \emph{Institute for Medical Engineering and Science, Massachusetts Institute of Technology, Cambridge, Massachusetts 02139, USA}}
\email{fferrett@mit.edu}
\affiliation{Dipartimento di Fisica, Universit\a`{a} Sapienza, 00185 Rome, Italy}
\affiliation{Istituto Sistemi Complessi, Consiglio Nazionale delle Ricerche, UOS Sapienza, 00185 Rome, Italy}
\author{Simon Grosse-Holz}
\affiliation{Department of Physics, Massachusetts Institute of Technology, Cambridge, Massachusetts 02139, USA}
\author{Caroline Holmes}
\affiliation{Department of Physics, Princeton University, Princeton, New Jersey 08544, USA}
\author{Jordan L Shivers}
\affiliation{Department of Chemical and Biomolecular Engineering, Rice University, Houston, Texas 77005, USA}
\affiliation{Center for Theoretical Biological Physics, Rice University, Houston, Texas 77030, USA}
\author{Irene Giardina}
\affiliation{Dipartimento di Fisica, Universit\a`{a} Sapienza, 00185 Rome, Italy}
\affiliation{Istituto Sistemi Complessi, Consiglio Nazionale delle Ricerche, UOS Sapienza, 00185 Rome, Italy}
\affiliation{INFN, Unit\a`{a} di Roma 1, 00185 Rome, Italy}
\author{Thierry Mora}
\affiliation{Laboratoire de Physique de l'\a'{E}cole Normale sup\a'{e}rieure (PSL University), CNRS, Sorbonne Universit\a'{e}, Universit\a'{e} de Paris, 75005 Paris, France}
\author{Aleksandra M Walczak}
\affiliation{Laboratoire de Physique de l'\a'{E}cole Normale sup\a'{e}rieure (PSL University), CNRS, Sorbonne Universit\a'{e}, Universit\a'{e} de Paris, 75005 Paris, France}

\date{\today}

\begin{abstract}
Flocking in $d=2$ is a genuine non-equilibrium phenomenon for which irreversibility is an essential ingredient. We study a class of minimal flocking models whose only source of irreversibility is self-propulsion and use the entropy production rate (EPR) to quantify the departure from equilibrium across their phase diagrams. The EPR is maximal in the vicinity of the order-disorder transition, where reshuffling of the interaction network is fast. 
We show that signatures of irreversibility come in the form of asymmetries in the steady state distribution of the flock's microstates.
They occur as consequences of the time reversal symmetry breaking in the considered self-propelled systems, independently of the interaction details. 
In the case of metric pairwise forces, they reduce to local asymmetries in the distribution of pairs of particles. 
This study suggests a possible use of pair asymmetries both to quantify the departure from equilibrium and to learn relevant information about aligning interaction potentials from data.

\end{abstract}


\maketitle

\section{Introduction}
Irreversibility is a distinguishing feature of active systems that enables remarkable collective phenomena not seen at equilibrium. 
Examples include motility-induced phase separation (MIPS), in which particles with strictly repulsive interactions segregate into dilute and dense phases spontaneously \cite{rev-MIPS-Cates-Tailleur}, and flocking, in which polar systems with short-ranged ferromagnetic interactions produce large-scale collective motion, even in $d=2$ \cite{annurev-Chate,revTTR}. 
These systems differ from their passive counterparts only in their constituents' self-propulsion, which can be viewed as a source of effective interactions, e.g. the effective attraction seen in MIPS \cite{effective-brader} or effective long-range alignment in flocking.

The key ingredient for the emergence of such collective phenomena is irreversibility. 
Self-propulsion can bring an active system out of equilibrium by injecting energy at the microscopic scale, even in the absence of alternative sources of irreversibility, like non-reciprocity of interactions or time delays \cite{Vitelli-nonreciprocal,nonmutual_only_ramaswamy,PRX-rev-Marchetti-Bowick-Nikta-Ramaswamy,Loos-heat-delay}. Yet, if the dynamics obeys detailed balance, even when particles are motile a Hohenberg-Mermin-Wagner-type theorem holds, preventing any spontaneous breakdown of the rotational symmetry of an equilibrium system in two dimensions \cite{Tasaki}. Irreversibility is then a necessary condition for the existence of 2D flocks exhibiting true long range order. 

The departure of active matter from equilibrium has become a topic of considerable interest in recent years \cite{howfar,Takatori-Brady,BoDabelowPRX,Pietzonka_lattice,Fakhri-Broedersz-irrev} and can be quantitatively addressed using various measures for the breakdown of detailed balance, like effective temperatures, violations of the fluctuation-dissipation theorem \cite{Doi-Mossa-Cugliandolo,Szamel-Teff,Levis_Teff,Maggi-Crisanti-Teff,DalCengio-fdt}, or entropy production \cite{PRX-mips-entropy,revAOUP,entropy-full-spinney}. Depending on the scale of interest, this is done employing either agent-based or field-theoretical descriptions. Most of the effort so far has focused on MIPS models, especially in the phase-separated state (see \cite{rev-mips-irrev} for a review). 
Much less work, mainly using fluctuating hydrodynamic models, has been done to systematically quantify irreversibility in polar dry active matter \cite{Cates-scoop,Ginelli-response}. 

Nonetheless, it has been noted that polar active matter has a stronger non-equilibrium character than apolar active matter. 
For instance, it is known that, in contrast to scalar field theories \cite{PRL-criticalIsing-Lee, GnanMaggi, CaballeroIsing}, activity is relevant (in the RG sense) for continuous flocking models, such that the critical properties of active polar systems are different from their passive counterparts \cite{Chen-Toner-Lee,TonerTu95,Toner-Tu-Ulm,swarms399}. 
Additionally, it has been shown that, in contrast to MIPS models where only repulsive forces are present, when nonconservative aligning torques are introduced
pressure is no longer a state function \cite{pressure,Solon-pressure}.

In this paper, we quantify the departure from equilibrium in a minimal agent-based model of flocking, akin to the standard Vicsek model \cite{Vicsek95}, where the system is described as an active ferromagnet composed of self-propelled spins. 
We focus on microscopic descriptions of flocks and study how they depart from equilibrium by means of the entropy production rate \cite{Maes1999,Maes2000}. We first use this quantity to measure the breakdown of the time reversal symmetry across the phase diagram, which we find is maximal in the vicinity of the order-disorder transition, where reshuffling of neighbors is most efficient. We relate irreversibility to asymmetries in the steady state distribution, the details of which depend on the interactions of the model.
We propose that signatures of irreversibility in the steady state distributions may then reveal relevant information about aligning interaction potentials in biological polar active systems.

\section{Stochastic thermodynamics of the Vicsek model}
We consider a continuous-time variant of the original two-dimensional Vicsek model, where the system is described as a set of interacting Active Brownian Particles (ABPs) performing rotational but not translational diffusion, which self-propel at a fixed speed $v_0$ and align to each other through short-ranged ferromagnetic interactions. 
The stochastic equation of motion reads:
\begin{flalign}
\label{eq-x}d\bold x_i &= v_0\bold e(\theta_i)dt,\\
\label{eq-th}d \theta_i &= F_i(\bold X,\boldsymbol \Theta)dt + \sqrt{2D}dW_i,
\end{flalign}
where $W_i(t)$ is a set of independent Wiener processes for $1\leq i\leq N$, $\bold e(\theta)=(\cos\theta,\sin\theta)$ is the orientation vector in $d=2$ and $F_i(\bold X,\boldsymbol \Theta)$ is the torque that reorients the $i$-th particle. 
Since we are interested in quantifying the effect of self-propulsion alone, we assume the torques are symmetric, ensuring that the action-reaction principle holds. In this way, the only source of irreversibility is particle motility and momentum is on average conserved.
We note that models typically employed to simulate flocks, like the standard Vicsek model \cite{Vicsek95,Peruani} or topological variants with fixed number of interacting neighbors \cite{Peruani_topo_crowded,topo,topo-balanced}, involve non-reciprocal interactions, generating additional irreversible phase space currents. Therefore we choose:
\begin{equation}
F_i(\bold X,\boldsymbol \Theta) =-\frac{\partial\mathcal H_{XY}\left(\boldsymbol\Theta;\bold n(\bold X)\right)}{\partial\theta_i},
\label{F-H}
\end{equation} 
where
\begin{equation}
\mathcal H_{XY}(\boldsymbol\Theta;\bold n) = -\frac{J}{2}\sum_{ij}n_{ij}\cos(\theta_i-\theta_j) 
\label{H-XY}
\end{equation}
is the Hamiltonian of an $XY$ model defined on a graph with a given adjacency matrix $\bold n$. 
Contrarily to the classical XY model, $\bold n$ is not constant but depends on time (through the $\bold X$ variables, Eq.~\eqref{eq-x}).

The dynamics described by Eqs.~\eqref{eq-x}--\eqref{eq-th} is Markovian in the phase space of the $N$-body system, whose general coordinate is $z = (\bold X, \boldsymbol \Theta)$, where $\bold X=(\bold x_1,\dots,\bold x_N)$ is the set of particle positions and $\boldsymbol\Theta=(\theta_1,\dots,\theta_N)$ is that of velocity directions. We recall the definition of the average entropy production as the Kullback-Leibler divergence between the path probability of a stochastic trajectory $z(t)$, for $0<t<\tau$, and its time-reversed $z^\dagger(t)$. 
For a general Markov process, where $p[z(t)] = p[z(t)|z(0)]p_0\left(z(0)\right)$, the average entropy production is decomposed as follows, in the absence of external driving:
\begin{equation}
\mathcal S(\tau) = D_{KL}\left(p[z(t)]||p[z^{\dagger}(t)]\right) = \mathcal S^{hk}(\tau) + \Delta \mathcal S_0.
\label{S-dec}
\end{equation}
We denote by $\mathcal S^{hk}=\langle\log p[z(t)|z(0)]\rangle - \langle\log p[z^{\dagger}(t)|z^{\dagger}(0)]\rangle$ the housekeeping entropy production \cite{Spinney-Ford}, with $\langle\cdot\rangle$ the average over the ensemble of (forward) trajectories, and by $\Delta \mathcal S_0=\langle\log p(z(0))\rangle - \langle\log p(z^{\dagger}(0))\rangle$ the variation in the Shannon entropy of the initial conditions of forward and backward paths. 
Let $\dot {\mathcal S}$ be the entropy production \emph{rate} (EPR), defined from \eqref{S-dec} as: 
\begin{equation}
\dot {\mathcal S} = \lim_{\tau\to\infty}\mathcal S(\tau)/\tau.
\end{equation}

In order to compute this quantity for the process in Eqs.~\eqref{eq-x}--\eqref{eq-th}, we need to  identify the parity under time reversal (T) of all the state variable coordinates. Positions are time reversal symmetric, i.e. ${\bold x}(t)\mapsto  {\bold x}^\dagger(t)={\bold x}(\tau-t)$, while we assume $\bold e(\theta)$ is time reversal antisymmetric, ${\bold e\left(\theta(t)\right)}\mapsto  {\bold e\left(\theta(t)\right)}^\dagger =-{\bold e(\theta(\tau-t))}$, so that the angular degrees of freedom acquire a global $\pi$ offset: $\theta^{\dagger}(t)=\theta(\tau-t)+\pi$. Let us notice that, in the absence of a translational diffusion term in Eq.~\eqref{eq-x}, this choice for the parity of $v_0\bold e(\theta)$ --- typically identified as the \emph{self-propulsion} --- is mandatory, since it must be interpreted as a physical velocity. Adopting, on the contrary, a T-even prescription would lead to a diverging EPR, due to the deterministic nature of Eq.~\eqref{eq-x} (see App.~\ref{app:A1}).

Provided that orientations are T-odd, the first equation is perfectly reversible and does not contribute to $\dot {\mathcal S}$. We can therefore eliminate the positional degrees of freedom and transform them into external parameters controlling the temporal evolution of $\bold n(t)= \bold n\left(\bold X(t)\right)$. In this way, the reference framework becomes that of a quasi-statically driven Langevin process, with state variables $\boldsymbol\Theta(t)$ and a set of time-varying parameters $\bold n(t)$ \cite{Seifert2008,Seifert2012}.

Introducing the Onsager-Machlup action \cite{OnsagerMachlup53} of the process into the formula of the housekeeping EPR and assuming stationarity, we obtain (see App.~\ref{app:A1}):
\begin{dmath}
	\dot{\mathcal S} = \frac{1}{D}\sum_{i}\langle\dot\theta_i(t)\circ F_i(t)\rangle = -\frac{J}{D}\sum_{ij}\langle	\dot\theta_i(t)\circ n_{ij}(t)\sin\left(\theta_i(t)-\theta_j(t)\right)\rangle,
	\label{epr-1}
\end{dmath}
where $\circ$ indicates the Stratonovich prescription.
According to the definition of stochastic heat of Langevin processes \cite{Sekimoto97}, Eq.~\eqref{epr-1} can also be interpreted as $\dot{\mathcal S}=D^{-1}\langle\,\dbar q/dt\rangle$, with $\,\dbar q$ the infinitesimal amount of heat dissipated into the medium (conventionally positive). The parameter $D$ corresponds to both the rotational diffusion coefficient of the ABP and the temperature of the heat bath that the active ferromagnet is in contact with.

Another expression for the EPR can be obtained by noticing that to write Eq.~\eqref{epr-1} we exploited the ergodicity of the dynamics. Reintroducing an explicit time average allows us to integrate by parts (see App.~\ref{app:A2}), yielding:
\begin{dmath}
	\dot{\mathcal S} = \lim_{\tau\to\infty}\frac{\langle-\mathcal H_{XY}\left(\boldsymbol\Theta(\tau);\bold n(\tau)\right) + \mathcal H_{XY}\left(\boldsymbol\Theta(0);\bold n(0)\right)\rangle}{D\tau} - \frac{J}{2D}\sum_{ij}\langle\dot n_{ij}(t)\circ\cos\left(\theta_i(t)-\theta_j(t)\right)\rangle. 
	\label{epr-2}
\end{dmath}
The first term in Eq.~\eqref{epr-2} is proportional to the rate of change of the system's internal energy, and vanishes under the assumption of stationarity. 
The second term corresponds to the work done on the system per unit time, divided by $D$. This interpretation is sound within the 
`external protocol' framework described above, where 
\begin{dmath}
\dbar w_{\mathrm{resh}} = -\frac{J}{2}\sum_{ij} dn_{ij}(t)\circ \cos\left(\theta_i(t)-\theta_j(t)\right)=\sum_{ij} dn_{ij}(t)\circ \frac{\partial \mathcal H_{XY}(\boldsymbol\Theta;\bold n)}{\partial n_{ij}}(t)
\end{dmath}
represents the infinitesimal work of fictitious reshuffling forces which rewire the adjacency matrix.
When the system is in a steady state, the average internal energy is constant, so the EPR includes only contributions from this irreversible work.

The fact that the dissipated heat coincides with the irreversible work of the external (fictitious) forces is typical of systems satisfying the local detailed balance condition \cite{Crooks}. 
The equality between Eq.~\eqref{epr-1} and the second term of Eq.~\eqref{epr-2} has been numerically verified (inset in Fig.~\ref{fig:phase-diag}.i for Model I). For the class of models described by Eqs.~\eqref{eq-x}--\eqref{eq-th} it follows from a combination of microscopic reversibility due to the Hamiltonian structure of the force (see App.~\ref{app:A2}), and a condition of stationarity (vanishing of the first term of Eq.~\eqref{epr-2}).

Generalizations of the EPR formulas to $d>2$ are obtained in App.~\ref{app:A3}.

\section{Numerical Results}

\begin{figure*}[!ht]
\includegraphics[width=.497\textwidth,trim={1.2cm 0.3cm 1.7cm 1 cm},clip]{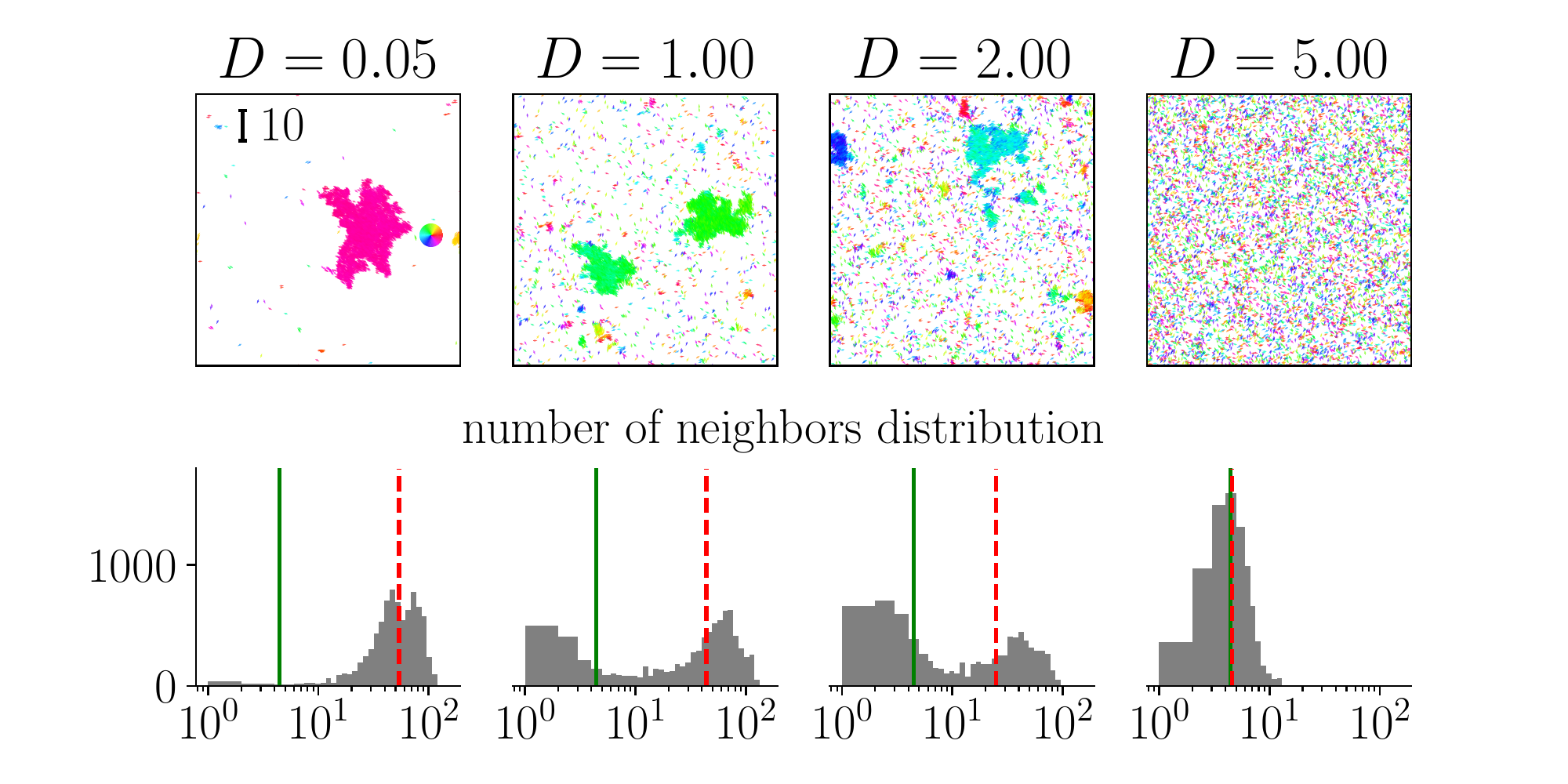}\llap{\parbox[b]{8.5 cm}{Model I\\\rule{0ex}{2in}}}\llap{\parbox[b]{17.2 cm}{\small{\textbf{a}}\\\rule{0ex}{1.9in}}}\llap{\parbox[b]{17.2 cm}{\small{\textbf{c}}\\\rule{0ex}{.85in}}}
\hfill
\includegraphics[width=.497\textwidth,trim={1cm 0.3cm 1.7cm 1 cm},clip]{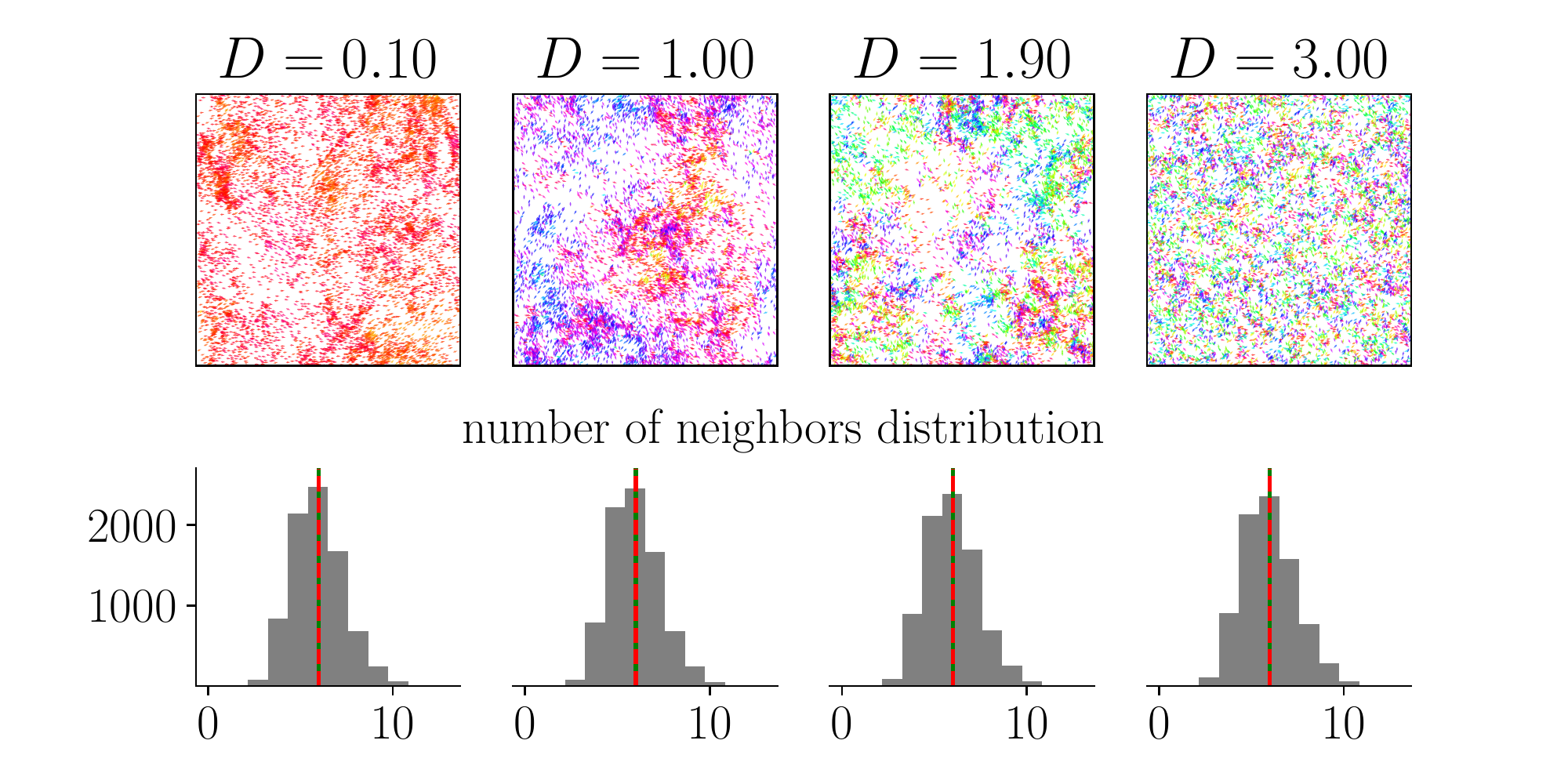}\llap{\parbox[b]{8.5 cm}{Model II\\\rule{0ex}{2in}}}\llap{\parbox[b]{17.2 cm}{\small{\textbf{b}}\\\rule{0ex}{1.9in}}}\llap{\parbox[b]{17.2 cm}{\small{\textbf{d}}\\\rule{0ex}{.85in}}}

\vspace{10pt}

\includegraphics[height=.2\textheight, trim={.6cm 0 .6cm 0},clip]{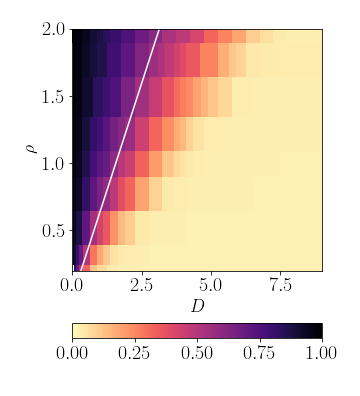}\llap{\parbox[b]{3.4 cm}{\small{Polarization}\\\rule{0ex}{1.8in}}}\llap{\parbox[b]{7.5 cm}{\small{\textbf{e}}\\\rule{0ex}{1.8in}}}
\includegraphics[height=.2\textheight, trim={.6cm 0 .6cm 0},clip]{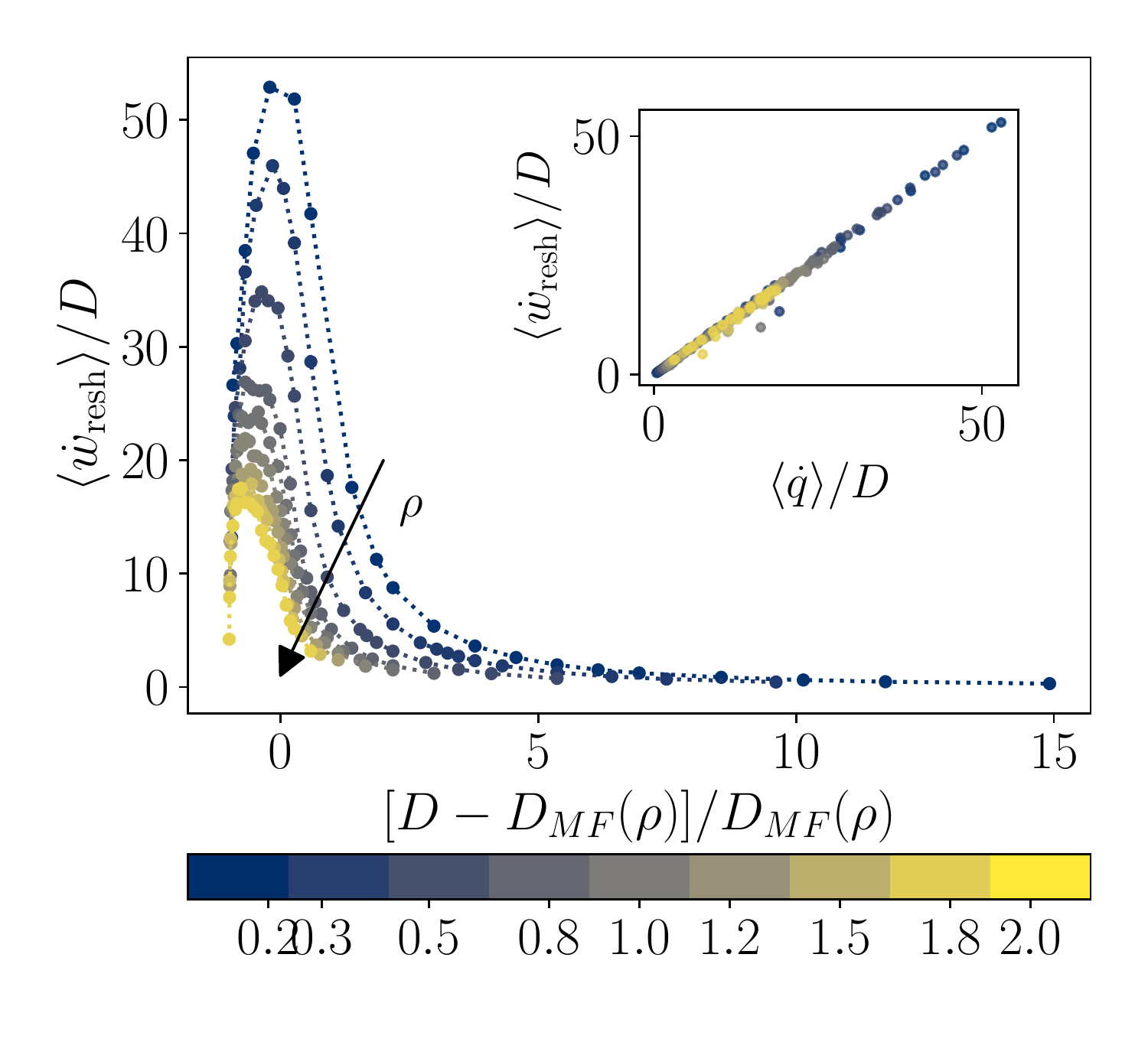}\llap{\parbox[b]{9.5 cm}{\small{\textbf{g}}\\\rule{0ex}{1.8in}}}
\hfill
\includegraphics[height=.2\textheight, trim={.6cm 0 .85cm 0},clip]{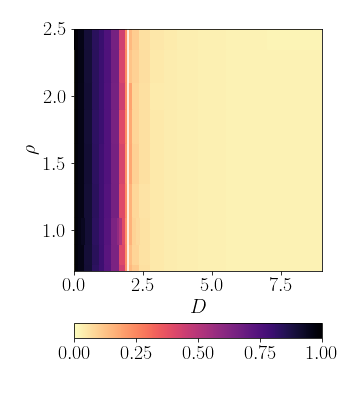}\llap{\parbox[b]{3.4 cm}{\small{Polarization}\\\rule{0ex}{1.8in}}}\llap{\parbox[b]{7 cm}{\small{\textbf{i}}\\\rule{0ex}{1.8in}}}
\includegraphics[height=.2\textheight, trim={.6cm 0 .8cm 0},clip]{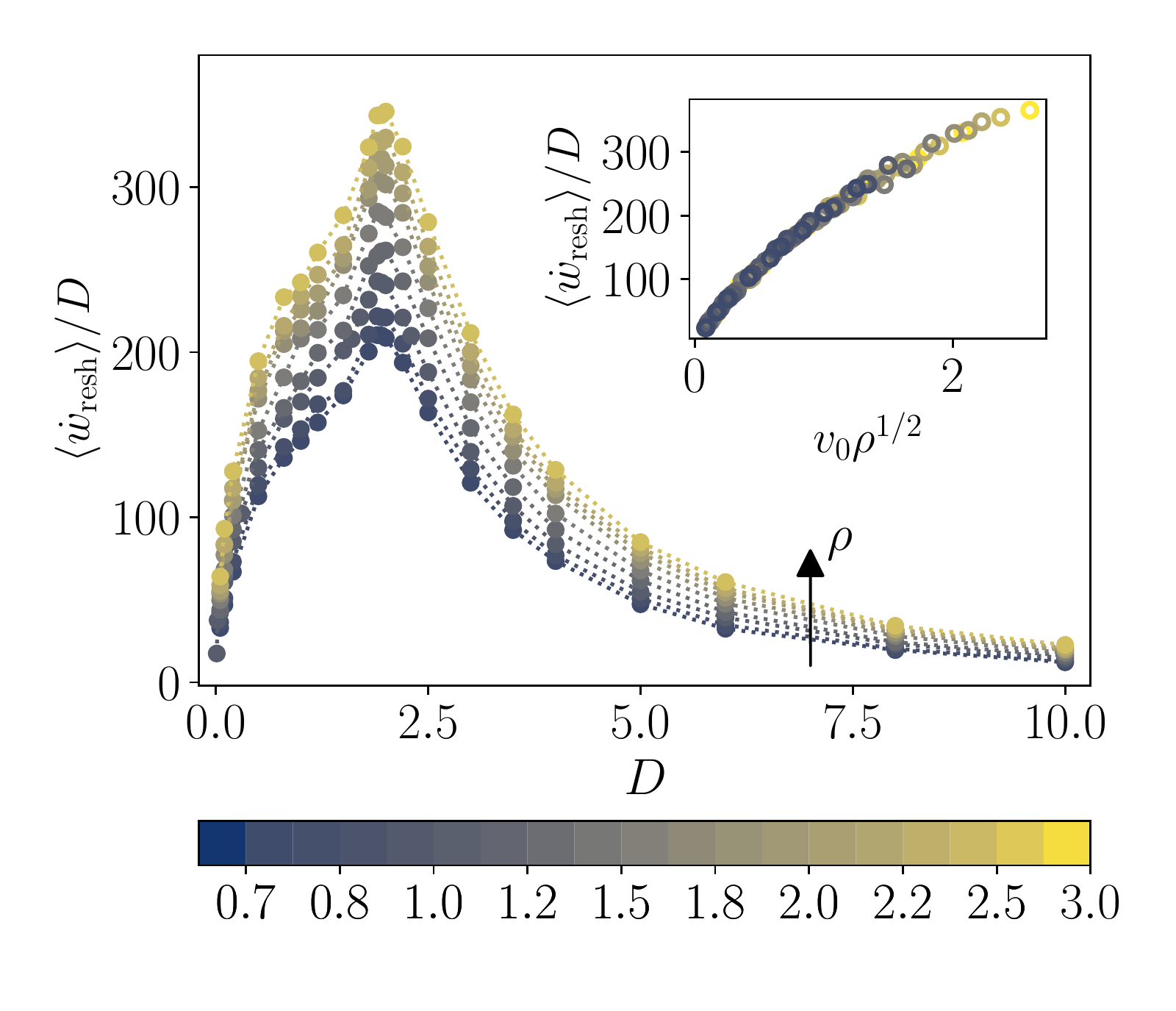}\llap{\parbox[b]{9.5 cm}{\small{\textbf{k}}\\\rule{0ex}{1.8in}}}

\includegraphics[height=.2\textheight, trim={.6cm 0 .6cm 0},clip,valign=t]{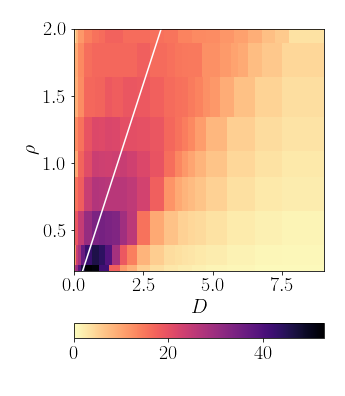}\llap{\parbox[b]{3.2 cm}{\small{EPR}\vspace{2pt}\rule{0ex}{0in}}}\llap{\parbox[b]{7 cm}{\small{\textbf{f}}\vspace{4pt}\rule{0ex}{.1in}}}
\includegraphics[height=.2\textheight, trim={.6cm 0 .8cm 0},clip,valign=t]{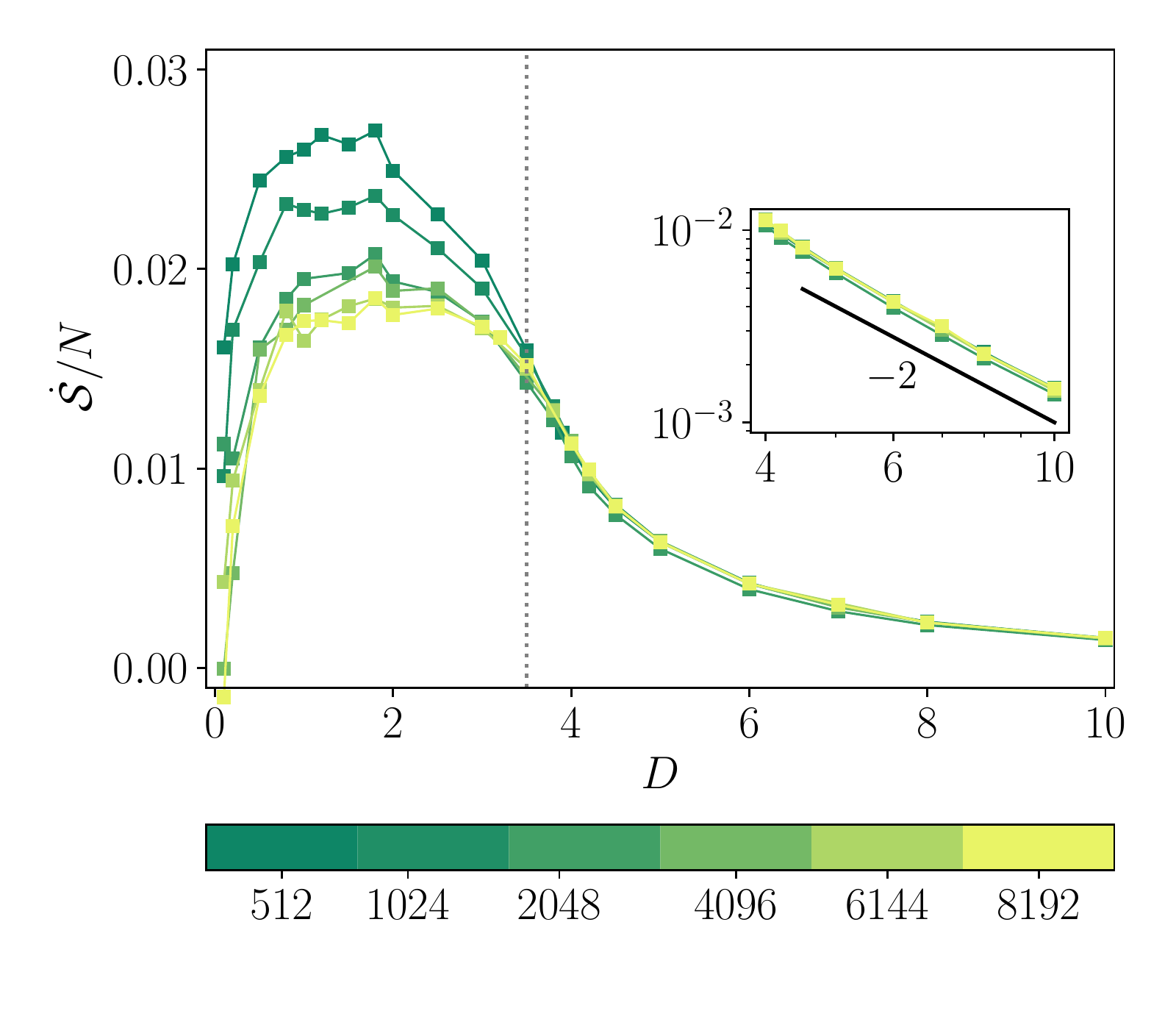}\llap{\parbox[b]{9.5 cm}{\small{\textbf{h}\vspace{4pt}}\rule{0ex}{.1in}}}
\hfill
\includegraphics[height=.2\textheight, trim={.75cm 0 .4cm 0},clip,valign=t]{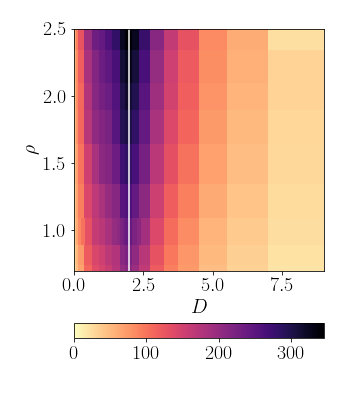}\llap{\parbox[b]{3.2 cm}{\small{EPR}\vspace{2pt}\rule{0ex}{0in}}}\llap{\parbox[b]{7 cm}{\small{\textbf{j}}\vspace{4pt}\rule{0ex}{.1in}}}
\includegraphics[height=.2\textheight, trim={.6cm 0 .4cm 0},clip,valign=t]{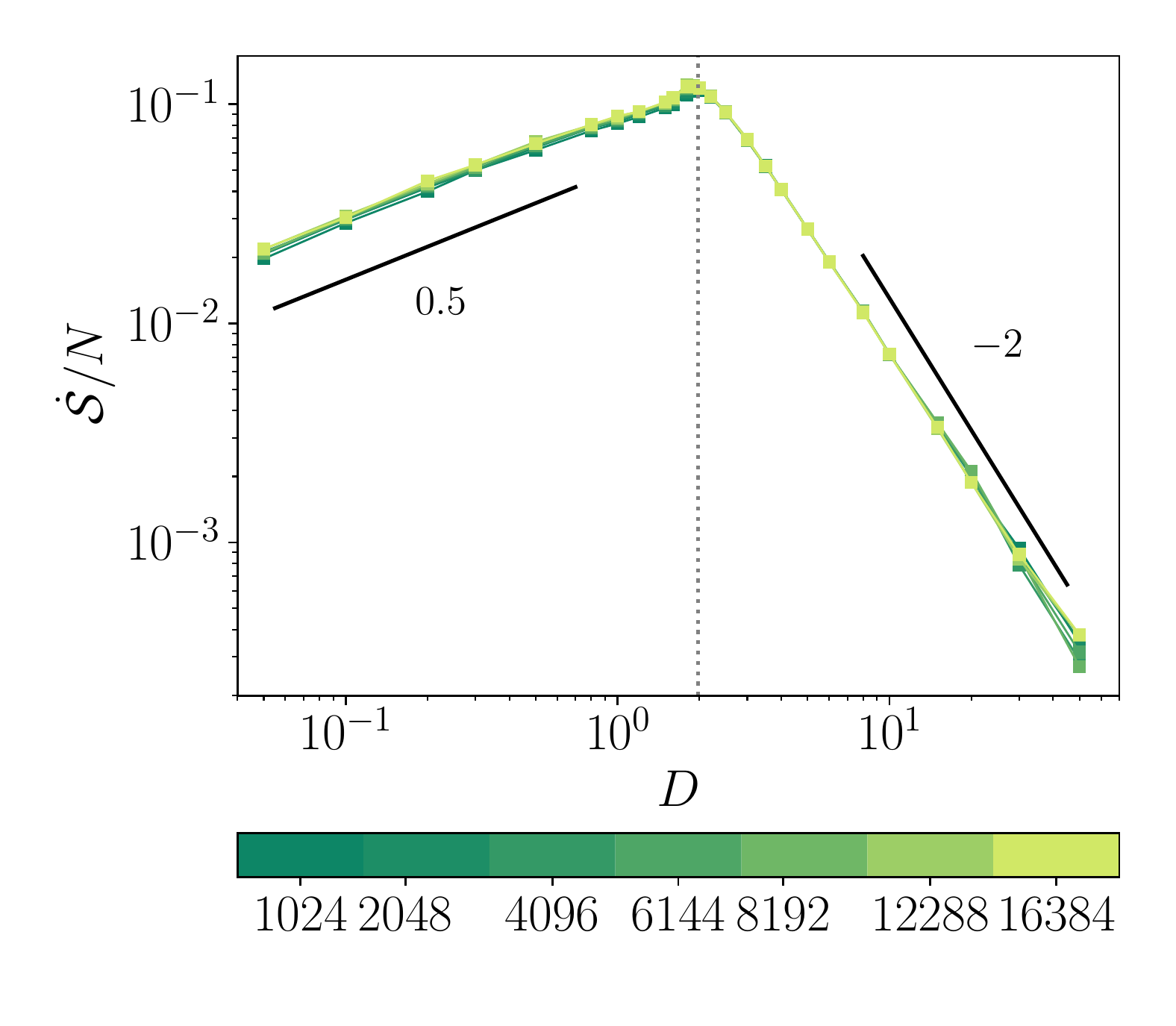}\llap{\parbox[b]{9.5 cm}{\small{\textbf{l}\vspace{4pt}}\rule{0ex}{.1in}}}
  
\caption{\textbf{a}--\textbf{b}: Typical configurations for Model I and Model II at different $D$ values. \textbf{c}--\textbf{d}: Distribution of the number of interacting neighbors associated to the configurations above. 
Bimodality at moderate to small $D$ values indicates phase coexistence in Model I. For Model II the distribution has an almost invariant shape, hence the amount of irreversible reshuffling of the interaction network cannot be deduced from its static connectivity alone. Red dashed line: mean number of neighbors; Green solid line: expected number of neighbors for a Poisson point patter on a torus ($\pi\rho \sqrt{2}R^2$ in the metric case \textbf{c}, 6 in \textbf{d}).
\textbf{e}: Polar order parameter (OP) of Model I. The white line is the mean field transition point \cite{Peruani}. 
\textbf{f}: EPR across the phase diagram of Model I. 
\textbf{g}: EPR curves as a function of the rescaled distance from the mean field transition point, $D_{MF}=J\rho\pi R^2/2$. A possible explanation for the remaining density dependence is provided in the main text. Inset: Equivalence of formulas \eqref{epr-1} and \eqref{epr-2} is verified, guaranteeing stationarity of the observed processes.
\textbf{h}: EPR per particle (Model I). Color code is associated to the system size $N$. Strong finite-size effects are evident for small $N$ in the symmetry-broken phase (where coexistence is realized), but $\dot {\mathcal S}/N$ seems intensive as $N$ is increased. The dashed vertical line represents the spinodal (extracted from the number of neighbors' distribution for $N=8192$).
\textbf{i}: OP of Model II.
\textbf{j}: EPR across the phase diagram of Model II. The white line represents the alleged critical point.
\textbf{k}: Density dependence of the EPR curves (discussed in the main text). Inset: irreversibility is governed by a single control parameter, $v_0\rho^{1/2}$, at fixed $D$.
\textbf{l}: EPR per particle for Model II (different colors for different $N$ values). All curves perfectly collapse on a single master curve, which converges to the two equilibrium limits as predicted.
Simulation parameters: (Model I) $R=1$, $J=1$, $v_0=0.5$; when not explicitly indicated $N=1024$, $\rho=1$. (Model II) $J=1$, $v_0=0.5$; when not explicitly indicated $N=2048$.
}
\label{fig:phase-diag}
\end{figure*}

Let us now parametrize the connectivity matrix $\bold n(\bold X)$ and use the formulas above to compute the EPR from numerical simulations of the model. We implement two variants of the Langevin-Vicsek model in Eq.~\eqref{eq-x}--\eqref{eq-th} with short-range ferromagnetic interactions.
In the first one (Model I) we model a metric pairwise alignment with 
\begin{equation}
n_{ij}^{(\mathrm I)}(\bold X) = \Theta(R-|\bold x_i-\bold x_j|),
\end{equation}
in which $\Theta$ is the Heaviside step function. 
The second variant (Model II) implements a topological multi-particle interaction, with
\begin{equation}
n_{ij}^{(\mathrm{II})}(\bold X) = 
\begin{cases}
1 \quad \mathrm{if\ } i,j\mathrm{ \ Voronoi\ neighbors} \\
0\quad \mathrm{otherwise}.
\end{cases}
\end{equation}
Both models are known to exhibit a phase transition from a disordered isotropic phase to a polar ordered phase, but with different phenomenology. Model I undergoes a first order phase transition, where sharp phase coexistence is realized in a wide portion of the ordered phase (see Fig.~\ref{fig:phase-diag}.a,c and \cite{disc-Greg-Chate}). On the contrary, spatial heterogeneities are largely suppressed in Model II, and the transition seems to be of second order at the considered system sizes (Fig.~\ref{fig:phase-diag}.b,d). For a discussion on the nature of the transition of active models with metric-free interactions see \cite{Vicsek-topo,BGL-topo,Tailleur-topo}). 

Phase diagrams for the modulus of the polar order parameter $\Phi=\frac{1}{N}\left\lvert\sum_{i=1}^N \bold e(\theta_i)\right\rvert$ and for the EPR are plotted in Figs.~\ref{fig:phase-diag}.e--l. We observe that, as control parameters are varied, the two models depart from equilibrium with roughly the same qualitative behavior: the entropy production rate peaks at intermediate $D$ values, while it vanishes as $D\to\infty$ or $D\to0$. 
A heuristic explanation for this behavior is readily provided if we recall that in the considered class of models non-equilibrium effects are entirely due to the rewiring of the interaction network. 

The existence of the first equilibrium limit is not surprising \cite{MarchettiEPR}: at $D\to\infty$ the system behaves as an ideal gas of free ABPs. The existence of a second equilibrium limit at $D\to0$ is less trivial, since it requires irreversible reshuffling to occur on time scales that diverge faster than $D^{-1}=\tau_P$ (persistence time). The reference model is a perfectly ordered passive ferromagnet, where reshuffling is suppressed by the strong alignment of particle velocities. This second limit justifies the use of approximate equilibrium descriptions for strongly interacting biological systems operating in this phase \cite{Mora:2016aa}.

A simple argument can be made to predict a power-law decay of the EPR in the two equilibrium limits, as detailed in App.~\ref{sec:scaling}. By approximating the amplitude of correlations and reshuffling times in these two limits, we deduce from Eq.~\eqref{epr-2} that the expected EPR scaling at high $D$ is $\dot{ \mathcal S} \sim D^{-2}$, whereas at low $D$ the EPR must scale as $\dot {\mathcal S}\sim D^{1/2}$.  
Figs.~\ref{fig:phase-diag}.l and \ref{fig:phase-diag}.h show the agreement between these predictions and numerical results for Model II (both regimes) and Model I (high $D$ regime only). 

Model-specific features in the EPR curves of Model I and Model II are also visible. For instance, the two variants of the Vicsek model show an opposite trend of the EPR with the average density of the system, $\rho=N/L^2$, as shown in Figs.~\ref{fig:phase-diag}.g and \ref{fig:phase-diag}.k. In Model I an increase in $\rho$ pushes the system towards an equilibrium-like mean field limit, by increasing the average number of neighbors and reducing the effect of reshuffling. Conversely, increasing $\rho$ in Model II causes an effective increase in the activity of the system. Since the average number of Voronoi neighbors does not depend on the density, but density modifies typical inter-particle distances, larger $\rho$ is equivalent to larger self-propulsion speed $v_0$. As shown in the inset of Fig.~\ref{fig:phase-diag}.k, the EPR of Model II is in fact governed by a single control parameter, $v_0\rho^{1/2}$.

The EPR curves of the two models also show a different shape and different sensitivity to finite size effects (Figs.~\ref{fig:phase-diag}.g vs. \ref{fig:phase-diag}.k and Figs.~\ref{fig:phase-diag}.j vs. \ref{fig:phase-diag}.n). This fact is due to the microscopic details of the reshuffling of the interaction network, which are fundamentally different in the two cases, as described in the next section. At low to intermediate values of the $D$ parameter, where the two curves are furthest from each other, Model I and Model II also differ in the features of their typical macroscopic configurations (most strikingly, the presence or absence of phase coexistence at the observed sizes). This fact corroborates the recently proposed idea that structure and dissipation are deeply interrelated in active matter \cite{SRho-Martiniani,Tociu-Vaik-Fodor}.

As a final note, the use of formulas \eqref{epr-1} and \eqref{epr-2} suggests that dissipation has a local origin in flocking models, but it does not seem to be spatially \emph{localized}, even if the system exhibits phase coexistence. In contrast to MIPS models, we could not observe a spatial segregation of the particles that relates to their dissipation \cite{PRX-mips-entropy,revAOUP,gonnella-work}. On the contrary, the heat or work contributions per single particle fluctuate in time, assuming both positive and negative values, with an amplitude comparable to the fluctuations of the mean EPR, rescaled by $\sqrt N$ (see Fig.~\ref{fig:non-localized} in App.~\ref{app:A6}).
However, since the EPR is a global quantity and its local decomposition is non-unique (we can always add a state function with a well-defined steady-state average value), we cannot exclude that different rewritings of $\dot {\mathcal S}$ could unveil alternative interesting interpretations.

\section{Signatures of irreversibility}

\begin{figure}[t]
\begin{minipage}[t]{.69\columnwidth}
\vspace{1pt}
\includegraphics[width=.99\textwidth]{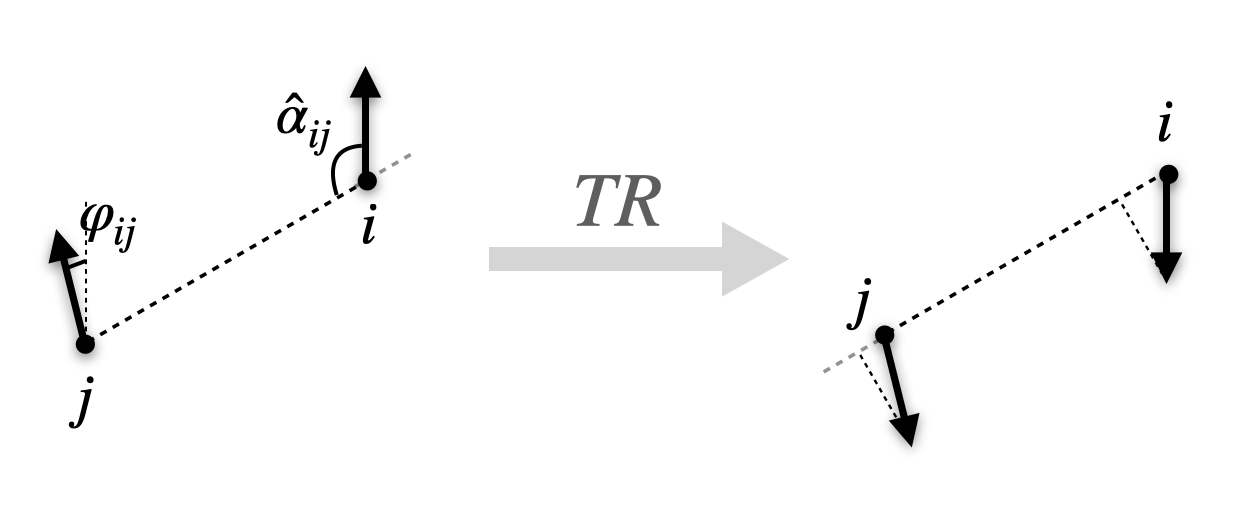}\llap{
  \parbox[b]{11cm}{\textbf{a}\\\rule{0ex}{.9in}}}
\vspace{.19cm}

\includegraphics[width=.9\textwidth,trim={.8cm 1cm 0.3cm 0},clip]{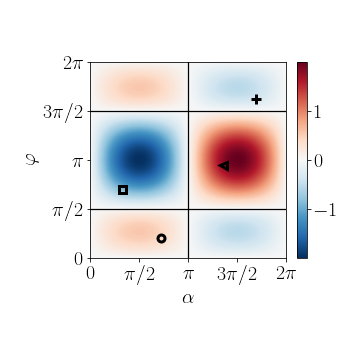}\llap{
  \parbox[b]{10.5cm}{\textbf{c}\hspace{10pt}\\\rule{0ex}{2.08in}}}\llap{
  \parbox[b]{4.5cm}{$\varepsilon(\alpha,\varphi)$\\\rule{0ex}{1.97in}}} 
\end{minipage}
\hfill
\begin{minipage}[t]{.295\columnwidth}
\vspace{0pt}
\includegraphics[width=.95\textwidth,trim={.5cm .5cm 0cm .6cm},clip]{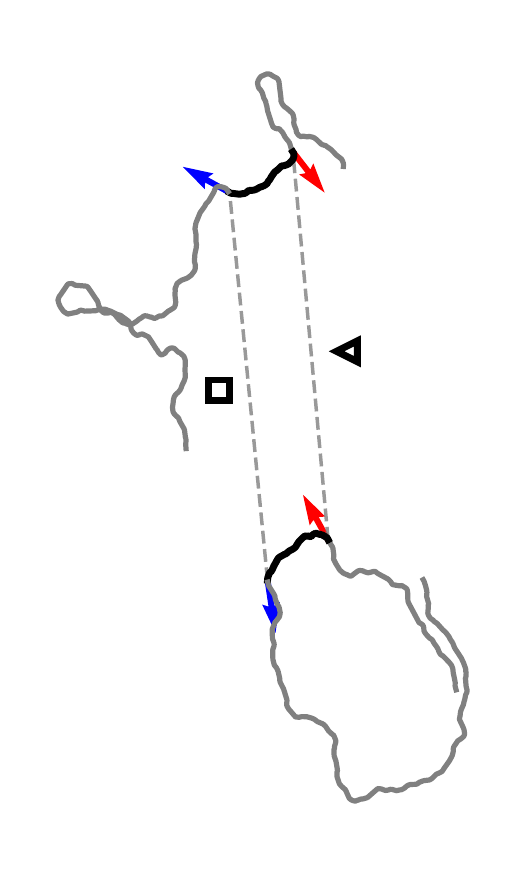}\llap{
  \parbox[b]{4.5cm}{\textbf{b.i}\\\rule{0ex}{1.35in}}}
\includegraphics[width=.63\textwidth,trim={1cm .6cm 1.5cm .6cm },clip]{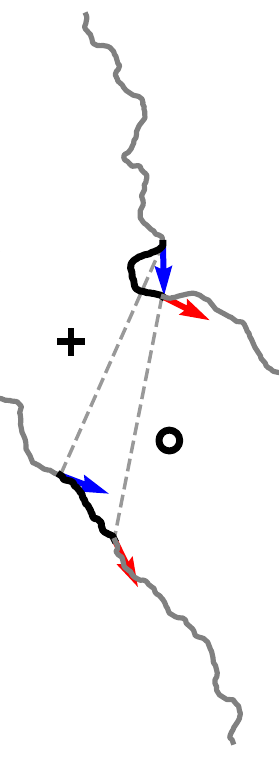}\llap{
  \parbox[b]{3.85cm}{\textbf{b.ii}\\\rule{0ex}{1.65in}}}
\end{minipage}
\caption{\textbf{a}: The time reversal (TR) operator acts on the system's variables by flipping velocities and keeping positions unchanged. Out of equilibrium, the two configurations cannot be equally probable.
\textbf{b}: Sample trajectories from a system of $N=1024$ particles ($D=3$, $J=1$, $v_0=0.5$, $\rho=1$, $\Phi\simeq0.21$) elucidating the dissipation mechanism in Model I. 
Black stretches indicate the portion of trajectory in which the two particles are at a distance smaller than the interaction radius; grey stretches those where particles are at a distance $r>R$ and do not interact.
In (\textbf{b.i}) strongly anti-aligned particles start interacting with an initial phase difference of $0.95\pi$ (triangle, converging red arrows, corresponding to a positive EPR contribution $\varepsilon_{in}\simeq1.76$). Due to rotational diffusion, they later leave each other's neighborhood at a new relative phase and displacement angle (square, blue arrows), corresponding to $\varepsilon_{out}\simeq-0.92$. In (\textbf{b.ii}) converging particles (cross) enter the interaction disk contributing to the EPR negatively ($\varepsilon_{in}=-0.34$). After the interaction, they diverge being more aligned than before (circle), so that the EPR contribution associated to this configuration surmounts the previous one: $\varepsilon_{out}=0.38$. 
\textbf{c}: Contour plot of Eq.~\eqref{epsilon}.}
\label{fig:scheme}
\end{figure}

Explicit expressions for the EPR of Vicsek-like models allow us to identify model-dependent signatures of irreversibility in the steady state distribution of a flock.
For the sake of simplicity, we focus on Model I and compute:
\begin{equation}
\dot n_{ij}^{(I)} = -\delta(R-|\bold x_i-\bold x_j|)\frac{(\bold x_i-\bold x_j)\cdot (\bold v_i-\bold v_j)}{|\bold x_i-\bold x_j|}.
\label{dotn}
\end{equation}
The rewiring of the connectivity matrix in Model I only depends on how mutual distances between pairs of particles evolve.
Let us insert Eq.~\eqref{dotn} into Eq.~\eqref{epr-2} and assume stationarity to rewrite the EPR as:
\begin{dmath}
\dot{\mathcal S} = \frac{Jv_0}{2D}\sum_{ij}\langle\cos\varphi_{ij} \left[\cos\hat\alpha_{ij} - \cos(\hat\alpha_{ij} - \varphi_{ij}) \right]\rangle_{|\bold x_i-\bold x_j|=R} 
\label{epr-asym}
\end{dmath}
where $\langle\cdot\rangle_{|\bold x_i-\bold x_j|=R}$ is the conditional average over pairs of particles at distance $R$. The variables $\varphi_{ij}$, $\hat\alpha_{ij}$ are angles parametrizing the mutual alignment and the relative angular position of the two particles, respectively (see Fig.~\ref{fig:scheme}):
\begin{equation}
\varphi_{ij} = (\theta_j -\theta_i) \bmod 2\pi\,,\quad \hat\alpha_{ij} = (\alpha^0_{ij}-\theta_i)\bmod 2\pi,
\label{angular}
\end{equation}
where $\alpha^0_{ij}$ is the angle indicating the direction of the displacement vector $\bold r_{ij} = \bold x_j -\bold x_i$ in a fixed reference frame, in which the $i$-th particle's orientation is $\theta_i$.
We can symmetrize Eq.~\eqref{epr-asym} by introducing $\alpha_{ij} = \hat\alpha_{ij} - \varphi_{ij}/2$ and rewrite 
\begin{equation}
\dot{\mathcal S} =  \frac{Jv_0}{2D}N^2g(R)\int_{[0,2\pi]^2} d\alpha d\varphi\, q(\alpha,\varphi)\varepsilon(\alpha,\varphi),
\label{epr-alpha-phi}
\end{equation}
where $g(r)=\frac{1}{N^2}\langle\sum_{ij}\delta \left(|\bold x_j - \bold x_i| - r\right)\rangle$ is the pair correlation function and
\begin{equation}
\varepsilon(\alpha,\varphi) = \cos\varphi\left[\cos\left(\alpha-\frac{\varphi}{2}\right) - \cos\left(\alpha+\frac{\varphi}{2}\right)\right]
\label{epsilon}
\end{equation}
is proportional to an EPR density per pair of particles. The quantity $q(\alpha,\varphi)$ is the (normalized) distribution of particle pairs at distance $R$:
\begin{dmath}
q(\alpha,\varphi) = \frac{\langle\sum_{ij}\delta(|\bold r_{ij}|-R) \delta(\hat\alpha_{ij}-\varphi_{ij}/2-\alpha)\delta(\varphi_{ij}-\varphi)\rangle}{N^2g(R)}.
\end{dmath}

The time reversal operator acts on the newly introduced angular variables in the following way: 
\begin{equation}
\varphi^{\dagger}(t)=\varphi(\tau-t),\quad \alpha^{\dagger}(t) = \alpha(\tau-t)+\pi.
\end{equation} 
Hence, breakdown of the time-reversal symmetry, $\dot {\mathcal S}\neq0$, implies from Eq.~\eqref{epr-alpha-phi} a symmetry breaking in the pair distribution: $q(\alpha,\varphi)\neq q(\alpha+\pi,\varphi)$. This means that two mirror configurations like those in Fig.~\ref{fig:scheme}.a cannot be equally probable. Specifically, since the EPR is non-negative, for aligned pairs ($\cos\varphi>0$) configurations with diverging particles are expected to be more probable than ones with converging particles. On the contrary, for anti-aligned pairs ($\cos\varphi<0$), configurations with colliding particles are expected to be more probable than ones with divaricating particles (cfr. Fig.~\ref{fig:scheme}.b). This scenario reflects the fact that birds leaving each other's neighborhood have been interacting in the past and are typically more aligned than those which have not interacted directly before.

\begin{figure}
\includegraphics[width=\columnwidth,trim={0.7cm 0 0.5cm 0},clip]{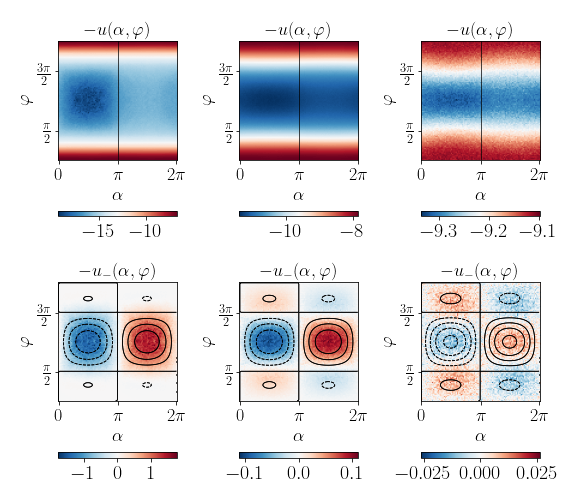}\llap{
 \parbox[b]{14.2cm}{$D=0.5$\\\rule{0ex}{3.1in}}}\llap{
 \parbox[b]{8.5cm}{$D=3.0$\\\rule{0ex}{3.1in}}}\llap{
 \parbox[b]{2.8cm}{$D=8.0$\\\rule{0ex}{3.1in}}}
\caption{First row: Reconstructed log-distributions of pairs of particles at distance $R$ from numerical simulations of Model I ($N=1024$, $v_0=0.5$, $\rho=1$, $J=1$). Second row: Antisymmetric part of the log-distributions: $-u_-(\alpha,\varphi) = u(\alpha+\pi,\varphi)-u(\alpha,\varphi)$. A positive correlation with $\varepsilon(\alpha,\varphi)$, shown in Fig.~\ref{fig:scheme}.c, is evident. Black lines represent the level curves of the fitted $\lambda\varepsilon(\alpha,\varphi)$ function, with $\lambda$ free parameter. }
\label{fig:pair}
\end{figure}

It is convenient to consider the logarithm of the particle pair distribution function, $u(\alpha,\varphi)=-\log q(\alpha,\varphi)$ and decompose it into T-symmetric and T-antisymmetric parts: $u_{\pm}(\alpha,\varphi) = \frac{1}{2}\left[u(\alpha,\varphi)\pm u(\alpha+\pi,\varphi)\right]$. The irreversibility condition now reads $u_-(\alpha,\varphi)\neq0$, while Eq.~\eqref{epr-alpha-phi} is rewritten as 
\begin{equation}
\dot{\mathcal S} \propto  \int_{[0,2\pi]^2} d\alpha d\varphi\,  e^{-u_+(\alpha,\varphi)}\sinh\left(-u_-(\alpha,\varphi)\right)\varepsilon(\alpha,\varphi).
\end{equation}
The positivity of the EPR is translated into a positive correlation between the log-distribution of particle pairs $-u_-(\alpha,\varphi)$ and the EPR density $\varepsilon(\alpha,\varphi)$. We show in Fig.~\ref{fig:pair} the numerically reconstructed function $-u(\alpha,\varphi)$, for different parameter values. 
In the disordered phase, the log-distribution looks almost T-symmetric, as the system is close to equilibrium (large $D$ in Fig.~\ref{fig:pair}, second row). 
In the ordered phase non-negligible asymmetries are visible in the reconstructed distributions: these are especially concentrated in the low probability region where particles are anti-aligned ($\varphi\sim \pi$).

\begin{figure}
\includegraphics[width=\columnwidth,trim={0.5cm .6cm 3.7cm .5cm},clip]{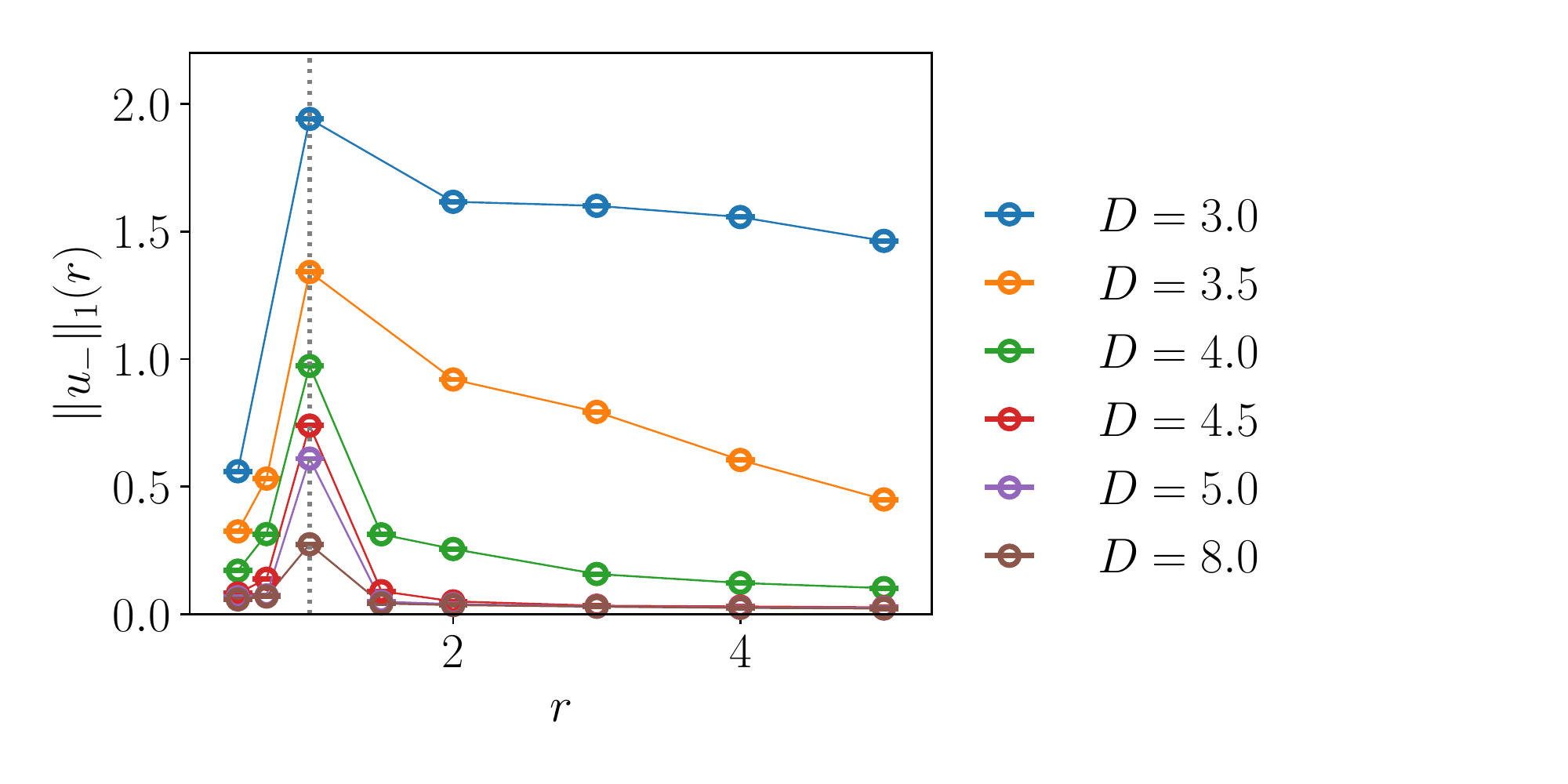}
\caption{Scalar measure for the degree of asymmetry of the log-distribution of particle pairs at distance $r$: $\lVert u_- \rVert_1(r)= \int d\alpha d\varphi \lvert u_-(\alpha,\varphi;r)\rvert $. Curves peak at $r = R$ (dashed line) for all the considered $D$ values.  Lowering the rotational diffusion coefficient, the decay to zero at large $r$ gets slower because of the presence of heterogeneous structures affecting the pair distribution. The trend of these curves is qualitatively reminiscent of $-\partial_rn(r) = \delta (r-R)$.  All data are collected form simulations of systems of $N=2048$ particles. }
\label{fig:peak}
\end{figure}

It is worth remarking that the discussed features are \emph{local}, as they are observed at the scale of the interaction radius $R$, which is much smaller than the system size or the typical size of polar clusters in the ordered phase. Asymmetries gradually disappear when we look at larger scales, as shown Fig.~\ref{fig:peak}. This property is not specific to Model I, but holds independently of the precise parametrization of $\bold n(\bold X)$, provided that it describes a pairwise short-ranged interaction. Interestingly, a similar asymmetric scenario has also been observed in \cite{Kursten-Ihle} in a variant of Model I with non-additive, non-pairwise interactions.

Model II is another example of a system with non-pairwise interactions, since $n_{ij}(\bold X)$ is not a simple function of the mutual distance between particles $i$ and $j$. As a consequence, we do not have a simple expression for $\dot n_{ij}$ to fill in Eq.~\eqref{epr-2}. Since the number of Voronoi cells must be conserved, reshuffling can occur only via the formation of $m$-fold vertices, with $m\geq4$. Such transitional configurations are realized when the particles form an $m$-sided polygon inscribed in a circle, so signatures of irreversibility must be sought for in $m$-particle densities (especially $m=4$, since other transitional configuration than the lowest order 4-fold vertex are unlikely to occur).

\section{Irreversibility-induced explicit symmetry breaking}
The fact that irreversibility constrains asymmetries in the two-body distribution (for Model I) is an example of a more general result concerning systems with T-odd state variable coordinates. Following the line of reasoning of \cite{DalCengio}, it can be shown that, given a stationary additive process $\bold z(t)$ described by a Langevin equation
\begin{equation}
\dot z^{\alpha} = A^{\alpha}(\bold z) + B^{\alpha\beta}\xi^\beta,
\label{SDE}
\end{equation}
whose coordinates have a definite parity under time reversal --- i.e. ${z^{\alpha}}^{\dagger}(t)=\epsilon^\alpha z^{\alpha}(\tau -t)$ with $\epsilon^\alpha=\pm1$ --- the condition $\dot {\mathcal S}>0$ implies that at least one of the two following statements is violated (see App.~\ref{app:B}):
\begin{flalign}
\label{cond1}A_{irr}^{\alpha}(\bold z) + D^{\alpha\beta}\partial_{\beta} \phi_+(\bold z) &= 0\ \ \forall \alpha,\\
\label{cond2}\partial_\alpha\phi_-(\bold z)&=0 \ \ \forall \alpha.
\end{flalign}
Here $\bold D = \frac{1}{2}\bold B^{\top}\bold B$ is the diffusion matrix of process \eqref{SDE}; $\bold A_{rev}$ and $\bold A_{irr}$ indicate the common decomposition of the drift term $\bold A$ into a reversible and an irreversible part \cite{Spinney-Ford};  $\phi_{\pm}$ are the T-symmetric and T-antisymmetric parts of the quasi-potential $\phi=-\log\psi(\bold z)$, where $\psi(\bold z)$ is the non-equilibrium steady state (NESS) distribution of the system.

When the state variable only contains even coordinates with respect to time reversal, Eq.~\eqref{cond1} must be violated, since $\phi_-(\bold z)=0$. In contrast, when $\bold A_{rev}(\bold z)=0$, the stationarity condition, combined with irreversibility, implies $\phi_-(\bold z)\neq0$ (see App.~\ref{app:B}). 
The considered 2D flocking model corresponds to this second scenario, as better described in App.~\ref{app:B}, hence the NESS distribution must be asymmetric under time reversal: 
\begin{equation}
\log\psi(\bold X,\boldsymbol\Theta)\neq\log\psi(\bold X,\boldsymbol\Theta+\pi).
\label{T-ESB}
\end{equation}
The discrete T symmetry which is here explicitly broken can be viewed as a $\pi$-rotation in the velocity subspace $\bold V = (\bold v_1,\dots \bold v_N)$, where $\bold v_k = e^{i\theta_k}$.
Therefore, a continuous rotational symmetry in the velocity subspace cannot hold; we can think of this fact as an irreversibility-induced explicit breakdown.

However, a different symmetry is preserved: the Fokker-Planck operator and the NESS distribution are invariant if arbitrary identical rotations are performed on both the external space of positions and the internal space of the polar order parameter:
\begin{equation}
\psi(e^{i{\theta_0}}\bold X,\boldsymbol\Theta + \theta_0)  = \psi(\bold X,\boldsymbol\Theta) \quad \forall \theta_0 \in \mathbb R. 
\label{SSB}
\end{equation}
As a consequence of Eq.~\eqref{SSB}, the marginalized pdf $\psi(\boldsymbol \Theta) = \int_{[0,L]^{2N}} d\bold X \psi(\bold X,\boldsymbol \Theta)$ is rotationally invariant, even when the dynamics of the polar active system is irreversible. Similarly, the marginalized distribution of positions, $\psi(\bold X) = \int_{[0,2\pi]^N} d\boldsymbol\Theta \psi(\bold X, \boldsymbol\Theta)$, is invariant under rotations.

In conclusion, the above discussion does not compel explicit symmetry breaking in $\psi(\boldsymbol\Theta)$, but underlines the lack of continuous symmetry under velocity rotations in the coupled $(\bold X,\bold V)$ space, which appears as a hypotesis in known Hohenberg-Mermin-Wagner theorems (including \cite{Tasaki}). In our case, this symmetry is only recovered at the equilibrium limits, where the absence of reshuffling disentangles positions and velocities.

\section{Conclusion}
Real life flocks and flocking models are often given as examples of strongly out of equilibrium systems. However, analysis of real flocks has showed that they can
function close to equilibrium if the self-propulsion leads the interaction network to rearrange on slow timescales compared to the local orientational dynamics \cite{Mora:2016aa}. 
Motivated by the observation that self-propulsion and irreversibility are not always synonymous and that the former alone is not a sufficient condition to explain the spontaneous emergence of collective behavior in polar active matter \cite{Tasaki}, we measured how flocks depart from equilibrium across their phase diagram. We employed the entropy production rate as the natural quantifier for the breakdown of detailed balance and exploited its positiveness to identify signatures of irreversibility in minimal agent-based flocking models.

While general statements can be deduced from the parity of the state variables alone, the way such signatures are manifested conveys model-specific information about alignment interactions in the flock. For the considered class of Vicsek-like models, violations of the time reversal symmetry are indeed due to the interplay between self-propulsion and the (otherwise equilibrium-like) alignment interactions. Self-propulsion makes particles motile, but interaction is required to probe the effect of motility, through the rewiring of the interaction network.

It is worth noting that rewiring could in principle be reversible, if it occurs in a symmetric way. However, the way information is transferred from velocity to positional degrees of freedom prevents this from happening. 
A better understanding of the thermodynamic and information-theoretical meaning of the rewiring of the interaction network may help understand not only the microscopic origin of the symmetry breaking out of equilibrium, but also how to control the robustness of the ordered phase in flocking systems.

Since irreversibility consists in a symmetry breaking (with respect to the time reversal transformation), it is not surprising that signatures of the out-of-equilibrium nature of the dynamics can be detected in asymmetries of the steady state distribution of the system. When the collective dynamics emerges from reciprocal pairwise interactions, such signatures of irreversibility can be detected in the pair distribution, at the scale of the interaction radius. Asymmetries are washed away on larger scales, if the system is sufficiently homogeneous.

Our analysis suggests that irreversibility-related features can be exploited as a tool to infer relevant information about microscopic interaction mechanisms in active polar systems from the experimental data.
Numerical simulations also show that the EPR is highest in the coexistence region of the metric additive Vicsek model, and peaks at the alleged critical point in the Voronoi Vicsek model, suggesting that the effects of irreversibility in polar active matter should be mostly visible in systems lying at the onset of collective motion. Candidate experimental models could be, for instance, moderate density actomyosin motility assays \cite{Frey-Bausch-actomyosin} or insect swarms \cite{swarms-scaling}.

Lastly, the numerically computed EPR exhibits a finite size scaling compatible with an extensive nature of the observable. It would be interesting to investigate how the EPR and the scaling we measure from agent-based microscopic models relates to the EPR of fluctuating coarse-grained theories and its scaling \cite{Caballero-stealth}.

\begin{acknowledgments}

The study was supported by the European Research Council Advanced Grant No. 785932,  the Italian Ministry of Foreign Affairs and International Cooperation through the Adinmat project, PRIN2020 grant n. 2020PFCXPE, and the European Research Council COG 724208. JLS was supported by the National Science Foundation Division of Materials Research (Grants No. DMR-1826623) and the National Science Foundation Center for Theoretical Biological Physics (Grant No. PHY-2019745). This work started as a summer school project during the 2019 Boulder School for Condensed Matter Physics: Theoretical Biophysics. IG and FF acknowledge valuable discussions with G. Gonnella in Rome and Bari.

\end{acknowledgments}

\appendix

\section{Derivation of EPR formulas}\label{app:A}

\subsection{Heat dissipation}\label{app:A1}

The process in Eq.~\eqref{eq-x}-\eqref{eq-th} is Markovian, so the path probability associated to it reads:
\begin{equation}
	p[\bold X(t),\boldsymbol \Theta(t)] \propto \psi(\bold X(0),\boldsymbol\Theta(0))e^{-\mathcal A_I[\bold X(t),\boldsymbol \Theta(t)]}
	\label{path-prob}
\end{equation}
where $\psi(\bold X, \boldsymbol\Theta)$ is the steady-state distribution of the $N$-body system, and
\begin{widetext}
\begin{equation}
	\mathcal A_I[\bold X(t),\boldsymbol\Theta(t)] = \lim_{\epsilon\to0}\lim_{\Delta t\to0}\sum_{s=0}^{N_s-1} \sum_{i=1}^{N} \frac{1}{\Delta t}\left\{ \frac{1}{\epsilon^2}\left[\bold x_i^{s+1} - \bold x_i^s - v_0 \bold e(\theta_i^s)\Delta t\right]^2 + \frac{1}{4D} [\theta_i^{s+1}-\theta_i^{s}+J\sum_j n_{ij}^s\sin(\theta_i^s-\theta_j^s)\Delta t]^2\right\}
	\label{A-OM}
\end{equation}
\end{widetext}
is the Onsager-Machlup (OM) action \cite{OnsagerMachlup53} (in the It\^o prescription --- any other $\alpha$-type prescription would lead to the same final expression for the EPR, since the process is additive). $N_s$ indicate the number of steps of amplitude $\Delta t$ in which the trajectory is discretized, and the limit is taken while keeping $N_s\Delta t=\tau$. 
The presence of the $\epsilon$-dependent term in Eq.~\eqref{A-OM} is due to the introduction of a translational diffusion term in Eq.~\eqref{eq-x} to regularize the $\delta(\dot{\bold x}_i-v_0\hat{\bold e}(\theta_i))$ that would appear from Eq.~\eqref{path-prob}.

Assuming that the system is in a steady state, the variation in the Shannon entropy of the initial state distribution vanishes, $\Delta \mathcal S_0=\langle\log\psi\left(\bold X(0),\boldsymbol\Theta(0)\right) - \log\psi\left(\bold X^{\dagger}(0),\boldsymbol\Theta^{\dagger}(0)\right)\rangle_0=0$, and the entropy production of the irreversible process reduces to the \emph{housekeeping} entropy production \cite{Spinney-Ford}: $\mathcal S(\tau)=\mathcal S^{hk}(\tau) = \mathcal A_I[\bold X(t),\boldsymbol\Theta(t)] - \mathcal A_I[\bold X^{\dagger}(t),\boldsymbol\Theta^{\dagger}(t)]$, where $\bold X^{\dagger}(t)$ and $\boldsymbol\Theta^{\dagger}(t)$ indicate the time-reversed trajectories.

Using Eqs.~\eqref{A-OM} and the OM action of the time-reversed trajectory, we can computing the entropy production \emph{rate} and obtain a first formula:
\begin{dmath}
{\dot{\mathcal S} = \frac{\langle\dot q\rangle}{D} = -\frac{J}{D}\sum_{ij}\langle\dot\theta_i\circ n_{ij}\sin(\theta_i-\theta_j)\rangle} =\lim_{N_s\to\infty}  \lim_{\Delta t\to0}\frac{1}{N_s\Delta t}\sum_{i=1}^{N}\sum_{s=0}^{N_s-1}\langle\frac{\theta_i^{s+1} - \theta_i^{s}}{\Delta t}\cdot\frac{F_i^s + F_i^{s+1}}{2}\rangle,
\label{work-epr}
\end{dmath}
where $F_i^s = F_i(\bold X^s, \boldsymbol\Theta^s) = -J\sum_{j}n_{ij}(\bold X^s)\sin (\theta_i^s - \theta_j^s)$ is the torque acting on the $i$-th particle at time step $s$, and $\langle\cdot\rangle$ is the average over the ensemble of non-equilibrium stationary paths. Here the limit is taken in such a way that $N_s\Delta t\to\infty$. Eq.~\eqref{work-epr} coincides with the average stochastic heat \cite{Sekimoto97} dissipated by the system into a heat bath at temperature $D$ per unit time, divided by $D$ or, equivalently, to the work rate of the aligning torques.

\subsection{Local equilibrium}\label{app:A2}

An alternative formula to Eq.~\eqref{work-epr} is obtained by eliminating the spatial d.o.f.s and projecting the process to those phase space directions where the irreversible current has nonzero components. In this reduced space, the diffusion matrix is invertible and no regularization is needed.
The effect of the eliminated d.o.f.s is taken into account by recognizing them as parameters driving the system through a quasi-static transformation. Having rephrased the problem in this way, we can follow \cite{Crooks}, and proceed by discretizing the trajectories of the angular d.o.f.s and of the protocol parameters as follows:
\begin{flalign*}
\mathrm{direct:}\quad&\boldsymbol\Theta_0\overset{\bold X_{1}}{\longrightarrow}\boldsymbol\Theta_1\overset{\bold X_{2}}{\longrightarrow}\boldsymbol\Theta_2\ \dots \ \boldsymbol\Theta_{M-1}\overset{\bold X_{M}}{\longrightarrow}\boldsymbol\Theta_{M};\\
\mathrm{time-reversed:}\quad&\boldsymbol\Theta_0\overset{\bold X_{1}}{\longleftarrow}\boldsymbol\Theta_1\overset{\bold X_{2}}{\longleftarrow}\boldsymbol\Theta_2\ \dots \ \boldsymbol\Theta_{M-1}\overset{\bold X_{M}}{\longleftarrow}\boldsymbol\Theta_{M}. 
\end{flalign*}
Let us now assume the local detailed balance condition \cite{Crooks}:
\begin{equation}
	\frac{P_c(\boldsymbol\Theta_n;\bold X_{n+1})P(\boldsymbol\Theta_n\overset{\bold X_{n+1}}{\longrightarrow}\boldsymbol\Theta_{n+1})}{P_c(\boldsymbol\Theta_{n+1};\bold X_{n+1})P(\boldsymbol\Theta_{n+1}\overset{\bold X_{n+1}}{\longrightarrow}\boldsymbol\Theta_{n})}=1,
    \label{db}
\end{equation}
where
\begin{equation}
    P_{c}(\boldsymbol\Theta;\bold X) = \frac{1}{Z(\bold X)}e^{-\beta \mathcal H_{XY}(\boldsymbol\Theta;\bold n(\bold X))}
\end{equation}
is the equilibrium Boltzmann distribution of an $XY$ spin system on a fixed network with connectivity matrix $\bold n(\bold X)$ and $\beta = D^{-1}$. Using hypothesis \eqref{db} to compute the entropy production of the discretized Markov process,
\begin{dmath}
{\mathcal S} = \langle\log \frac{P_c(\boldsymbol\Theta_0;\bold X_0)}{P_c(\boldsymbol\Theta_M;\bold X_M)}+\log\frac{\prod_{n=0}^{M-1} P(\boldsymbol\Theta_n\overset{\bold X_{n+1}}{\longrightarrow}\boldsymbol\Theta_{n+1})}{\prod_{n=0}^{M-1} P(\boldsymbol\Theta_{n+1}\overset{\bold X_{n+1}}{\longrightarrow}\boldsymbol\Theta_{n})}\rangle = \langle \log\prod_{n=0}^{M-1}\frac{P_c(\boldsymbol\Theta_{n};\bold X_{n})}{P_c(\boldsymbol\Theta_n;\bold X_{n+1})}\rangle,
\end{dmath}
and taking the continuous limit yields for the EPR:
\begin{equation}
\dot{\mathcal S} = \beta \langle \sum_{i<j}\dot n_{ij} \circ\left[\frac{\partial\mathcal H_{XY}(\boldsymbol\Theta;\bold n)}{\partial n_{ij}} - \frac{\langle\partial\mathcal H_{XY}(\boldsymbol\Theta;\bold n)\rangle_{c,\bold n}}{\partial n_{ij}} \right]\rangle.
    \label{local-db}
\end{equation}
The same result is obtained by applying the standard rules of stochastic calculus to Eq.~\eqref{work-epr}. Recalling that this expression involves a time average (eliminated under the assumption of ergodicity), we can perform an integration by parts resulting into:
\begin{dmath}
\dot{\mathcal S}  = -\frac{J}{2D}\sum_{ij}\langle\dot n_{ij}\circ\cos(\theta_i-\theta_j)\rangle + \lim_{\tau\to\infty}\frac{\langle\mathcal H_{XY}(\tau) - \mathcal H_{XY}(0)\rangle}{\tau},
    \label{int-epr}
\end{dmath}
where $\langle\cdot\rangle_{c,\bold n}$ is the equilibrium over the canonical ensemble, for fixed $\bold n$.

Let us now assume that the system is in a steady state, where the average $XY$ Hamiltonian does not vary with time (or at most sublinearly). Eq.~\eqref{int-epr} then reduces to the rate of work which fictitious reshuffling forces make on the system:
\begin{equation}
	\dot{\mathcal S} = \frac{\langle\dot w_{\text{resh}}\rangle}{D} = -\frac{J}{2D}\sum_{ij}\langle\dot n_{ij}\circ\cos(\theta_i-\theta_j)\rangle.
    \label{resh-epr}
\end{equation}
Let us remark that these fictitious forces are not independent of the system's state: thanks to this crucial dependency between the external protocol and the state of the system we can have a nonzero EPR.

\subsection{Arbitrary dimension}\label{app:A3}

The generalization of the active $XY$ model to the $d>2$ case is the active $O(n)$ ferromagnet, whose dynamics is described by:
\begin{flalign}
\label{d-eq-1} d x_i^{\alpha} &= v_i^{\alpha} dt ,\\
\label{d-eq-2} d v_i^{\alpha} &= P_i^{\alpha\beta}\left(-J\sum_jn_{ij} v_j^{\beta} + v_0\xi_i^{\beta}\right),
\end{flalign}
where $\langle\xi_i^{\alpha}(t)\xi_i^{\beta}(t')\rangle = \delta_{ij}\delta_{\alpha\beta}\delta(t-t')$ and $P^{\alpha\beta}_i(t) = \delta^{\alpha\beta} - v_i^\alpha(t)v_i^\beta(t)/v_0^2$ is the orthogonal projector to $\bold v_i(t)$ that is required for normalization purposes. For $d>2$ the process is multiplicative: the prescription to integrate Eqs.~\eqref{d-eq-1}--\eqref{d-eq-2} is the Stratonovich one.

The corresponding generalization of the EPR formulas~\eqref{work-epr} and \eqref{resh-epr} to the $d$-dimensional case reads:
\begin{flalign}
\label{epr-v-1}\dot{\mathcal S}  &= \frac{\langle\,\dbar q\rangle}{dt} = \frac{J}{Dv_0^2}\sum_{ij}\langle\dot v_i^\alpha(t)\circ P^{\alpha\beta}_i(t)  n_{ij}(t)v_j^{\beta}(t)\rangle;\\
\label{epr-v-2}\dot{\mathcal S} &= \frac{\langle\,\dbar w\rangle}{dt} = -\frac{J}{2D v_0^2}\sum_{ij}\langle\dot n_{ij}(t)\circ v_i^{\alpha}(t) v_j^{\alpha}(t)\rangle.
\end{flalign}

\subsection{Equilibrium limits}\label{app:A4}

\begin{figure*}[t!]
\begin{minipage}[t][][b]{.36\textwidth}
\includegraphics[width=.45\textwidth,trim={3cm 3cm 0 0},clip]{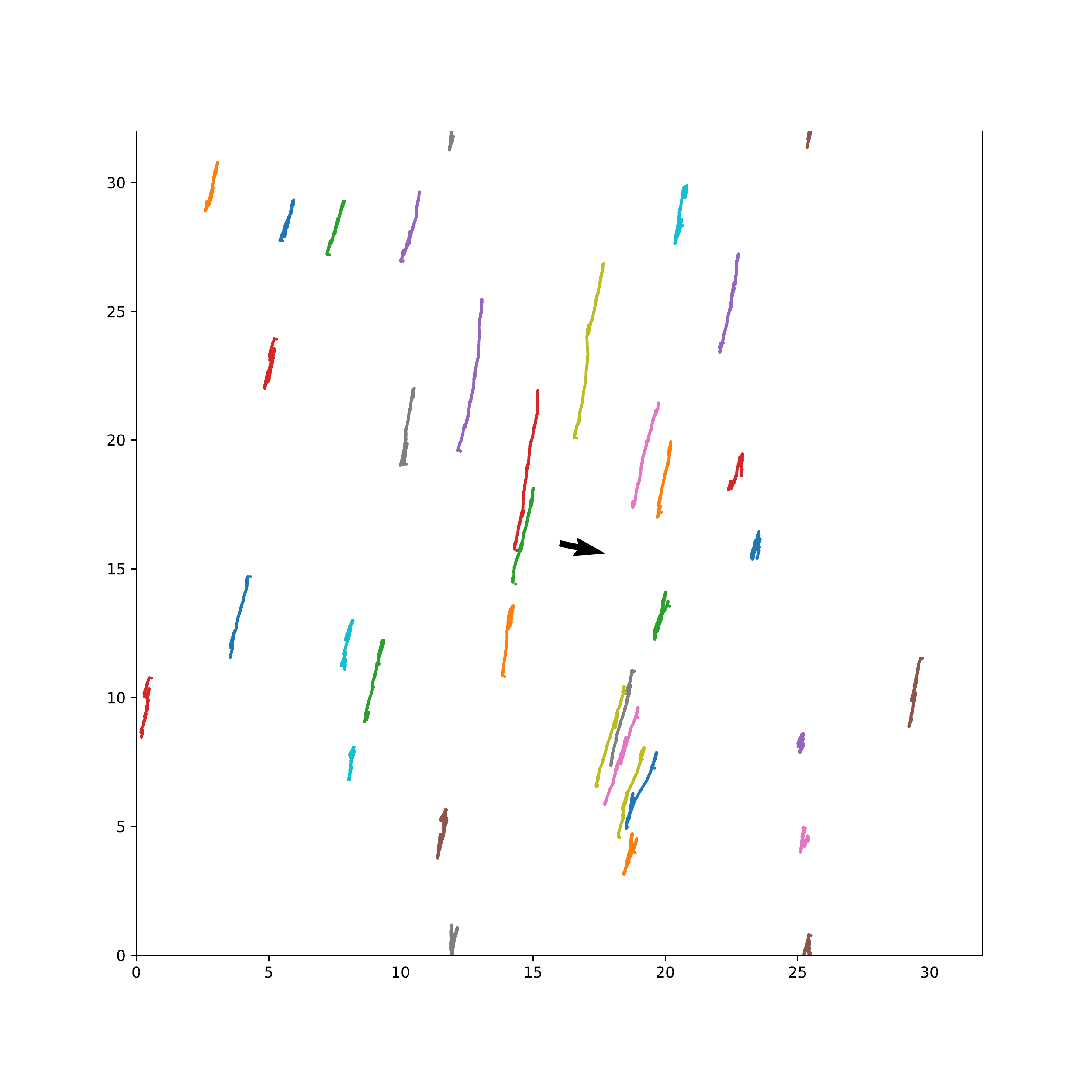}\llap{  \parbox[b]{6cm}{\textbf{\small{a}}\\\rule{0ex}{1.1in}}}\llap{
        \parbox[b]{3.2cm}{\scriptsize{$D=0.1$}\\\rule{0ex}{1.02in}}}
\includegraphics[width=.45\textwidth,trim={3cm 3cm 0 0},clip]{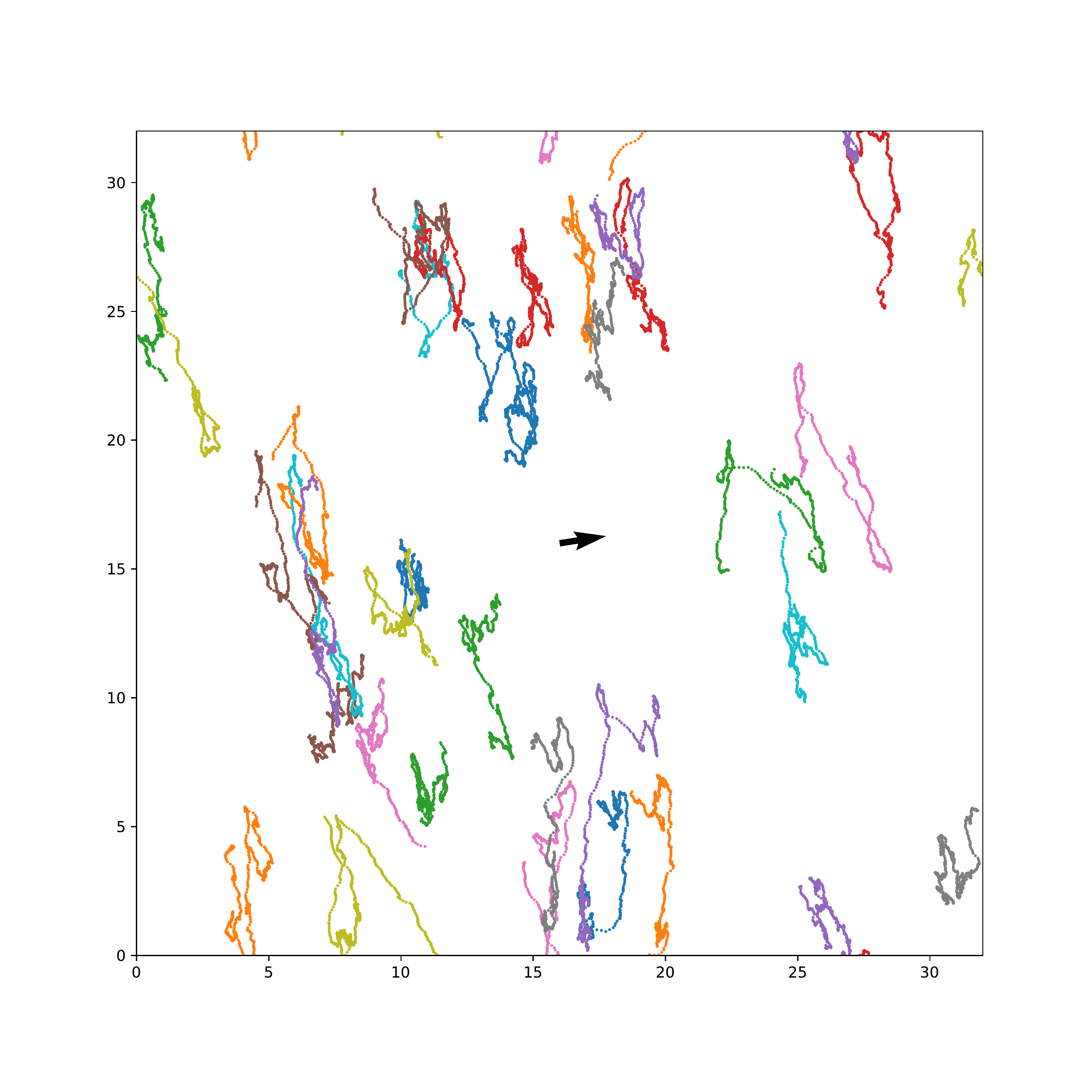}\llap{\parbox[b]{3.2cm}{\scriptsize{$D=1$}\\\rule{0ex}{1.02in}}}

\includegraphics[width=.45\textwidth,trim={3cm 3cm 0 0},clip]{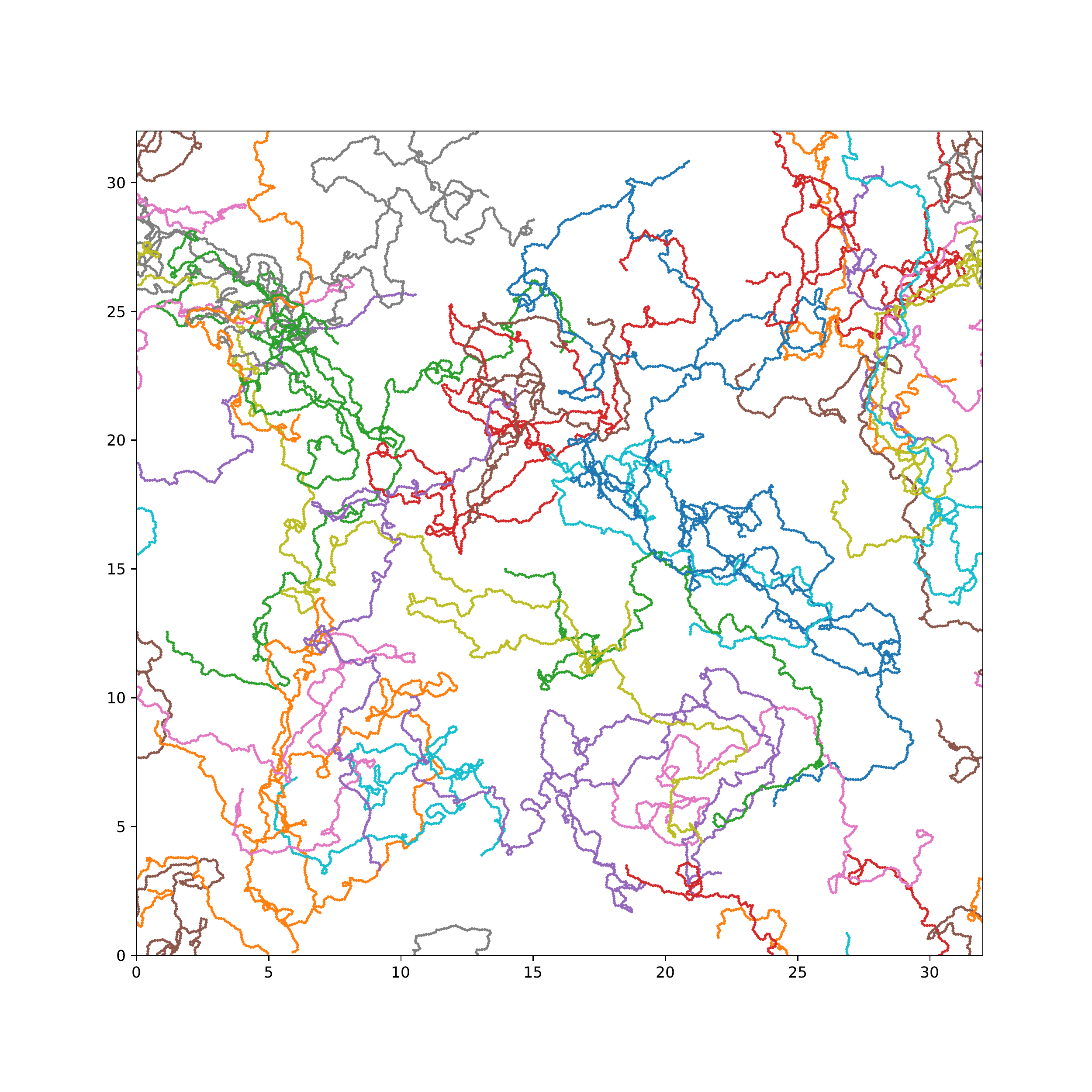}\llap{\parbox[b]{3.2cm}{\scriptsize{$D=2$}\\\rule{0ex}{1.02in}}}
\includegraphics[width=.45\textwidth,trim={3cm 3cm 0 0},clip]{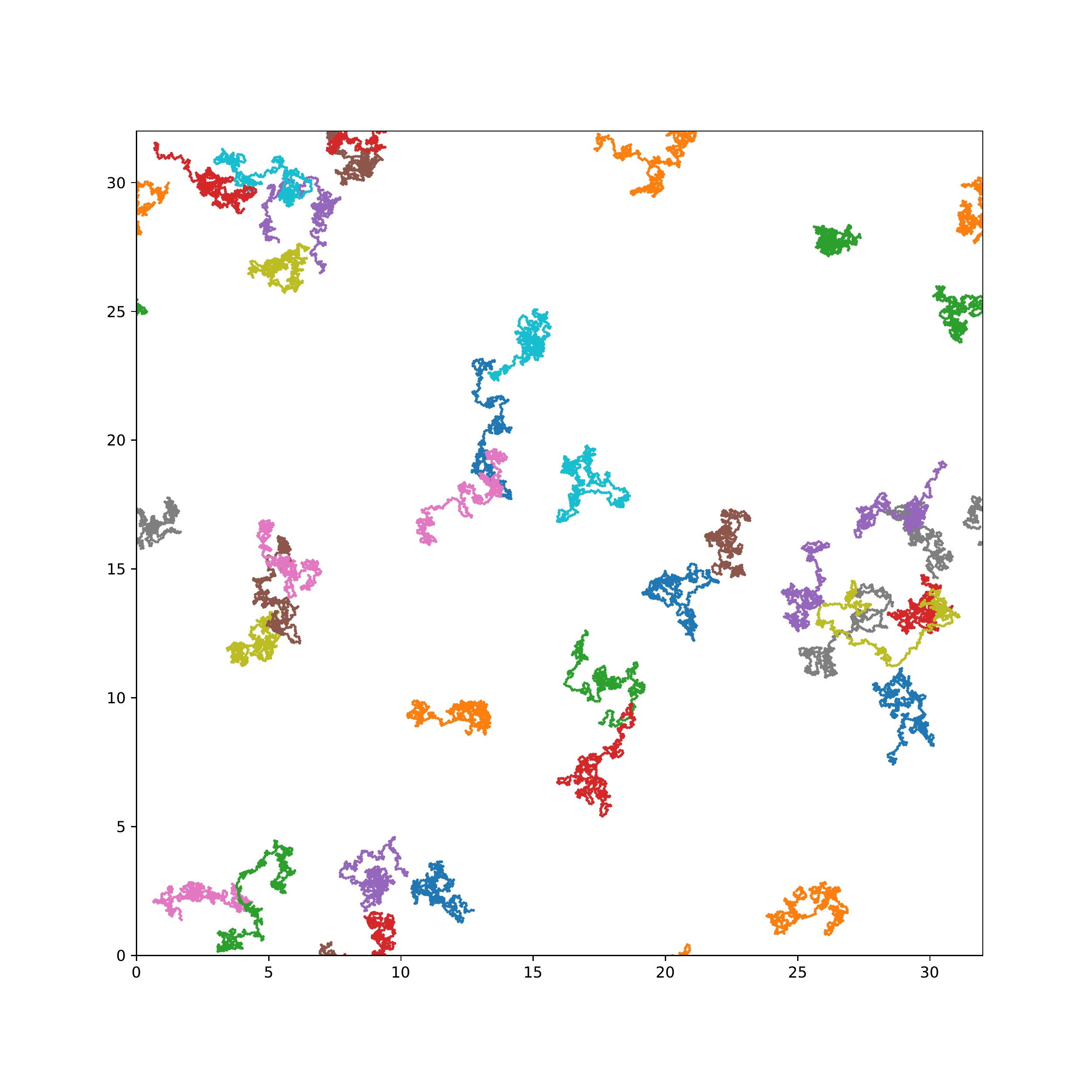}\llap{\parbox[b]{3.2cm}{\scriptsize{$D=8$}\\\rule{0ex}{1.02in}}}
\end{minipage}
\hfill
\begin{minipage}[t][][b]{.6\textwidth}
\vspace{.7cm}
\includegraphics[width=.6\textwidth,trim={ 0.7cm 0 1cm 0},clip]{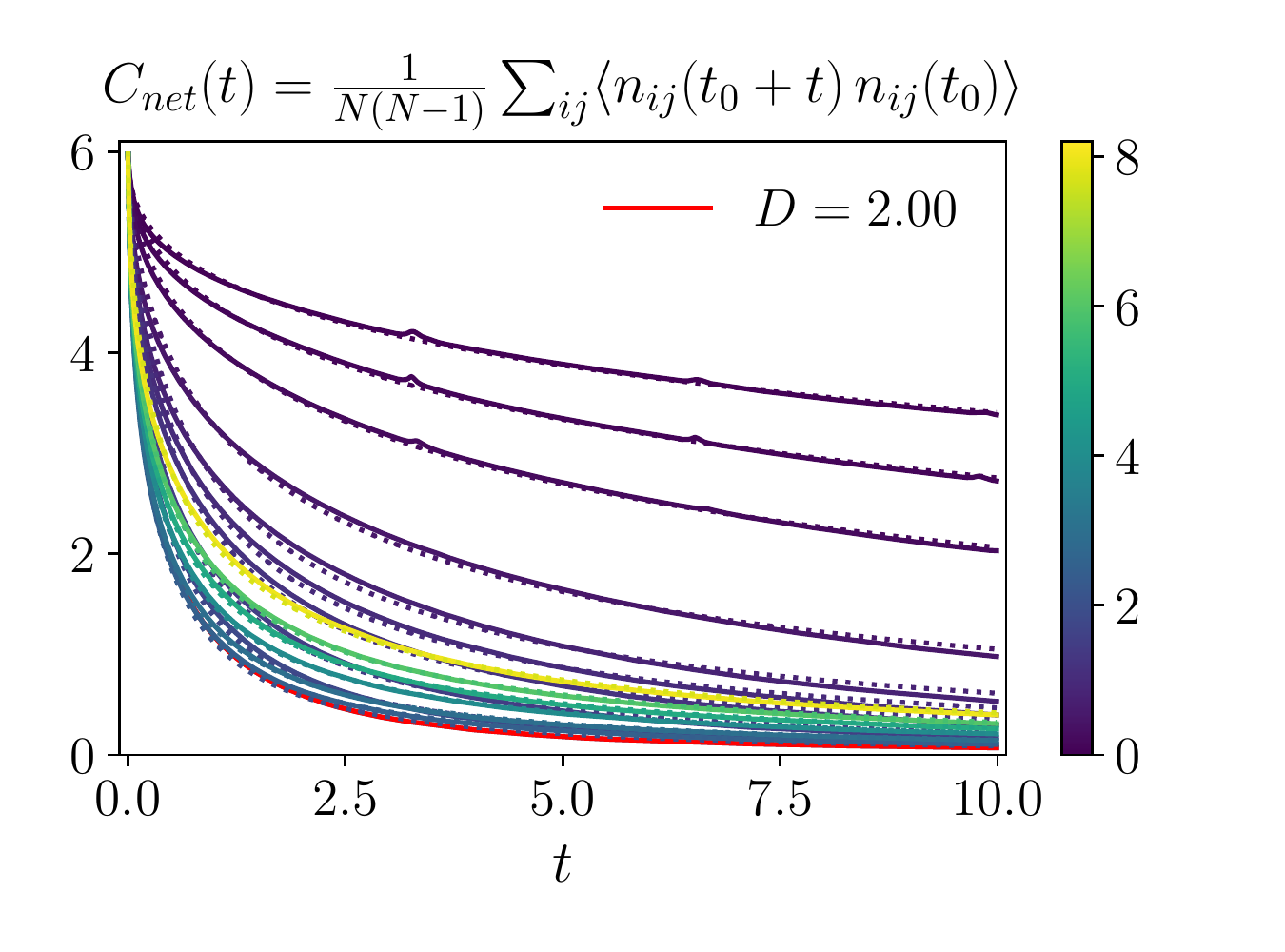}\llap{\parbox[b]{14cm}{\textbf{b}\\\rule{0ex}{2.in}}}	
    \hfill
    \includegraphics[width=.38\textwidth,trim={1cm 0cm .2cm .65cm },clip]{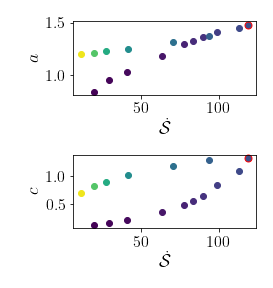}\llap{\parbox[b]{8cm}{\textbf{c}\\\rule{0ex}{1.9in}}}\llap{\parbox[b]{8cm}{\textbf{d}\\\rule{0ex}{.95in}}}
\end{minipage}
    \caption{ \textbf{a}: Subsample of trajectories of diffusing particles in different regimes. All the trajectories span the same time interval. In the symmetry broken phase the trajectories are shown in the reference frame of the center of mass; mutual diffusion occurs mainly in the transverse direction with respect to collective motion (black arrow). At $D=2$, where reshuffling is mostly efficient, the flock is disordered but the motion of the particles is still persistent. Persistence is reduced as the rotational diffusion coefficient $D$ is increased. In all cases, $N=1024$, $\rho_0=1$. \textbf{b}: Autocorrelation function of the adjacency matrix of the flock, $C_{\text{net}}(t)$, defined as in Eq.~\eqref{Cnet-def}. The color map refers to $D$ values. The maximum point of the EPR, $D=2$, is marked in red and corresponds to the curve with the fastest decay. Dashed lines are the fitted curves from Eq.~\eqref{fit-Cnet}.  \textbf{c}--\textbf{d}: Parametric plot of fitted parameters, $a$ and $c$, versus EPR. The figure shows a positive correlation both for the effective diffusion coefficient $c$ and exponent $a$. Close to the transition point (marked by the red dot) reshuffling is the most efficient and the system is in the farthest condition from an equilibrium one.}
    \label{fig:topo-diff}
\end{figure*}

The obtained EPR formulas reveal the existence of several equilibrium limits for the Vicsek model. Two of them obviously correspond to $v_0\to0$, where activity is suppressed and the model corresponds to an equilibrium $XY$ ferromagnet, and $J\to0$. 
From inspection of Eq.~\eqref{work-epr} a third equilibrium limit can be deduced, for $D\to\infty$, since torques and phase increments are bounded. In this limit, the interaction term in Eq.~\eqref{eq-th} becomes negligible, compared to the noise, and the system behaves as in the non-interacting equilibrium case. 
Because of the equivalence between Eqs.~\eqref{work-epr} and \eqref{resh-epr}, this means that reshuffling is suppressed as the rotational diffusion coefficient increases. Particles tend indeed to swirl around their positions, being stack in the vicinity of their own neighborhood for a long time (see Fig.\ref{fig:topo-diff}).

Another equilibrium limit, which is independent of the parametrization of $n_{ij}$, corresponds to $D\to0$. In the strongly polarized phase the system approaches the behavior of a passive ferromagnet in the co-moving reference frame. The relevant (asymmetric) contribution of reshuffling is suppressed faster than $D$ in this limit, resulting in an effective equilibrium behavior. An approximate argument for this fact is provided in \ref{sec:scaling}.

Two other nontrivial equilibrium limits can be deduced for the two variants of the Vicsek model considered in the main text, by varying the control parameter $\rho$. 
In the additive metric case (Model I) the $\rho\to\infty$ limit corresponds to an equilibrium limit, since the system approaches an effective mean field configuration.
In the momentum-conserving topological case (Model II) the opposite limit, $\rho\to0$, is an equilibrium one.

Enhancement of reshuffling close to the transition is witnessed by the decay of the reconstructed autocorrelation function of the connectivity matrix:
\begin{equation}
	C_{\text{net}}(t)=\frac{1}{N(N-1)}\sum_{ij}\langle n_{ij}(t_0+t)n_{ij}(t_0)\rangle_{t_0},
	\label{Cnet-def}
\end{equation}
where $\langle\cdot \rangle_{t_0}$ denotes a time average over multiple starting times. We show the quantity \eqref{Cnet-def} computed for Model II in Fig.~\ref{fig:topo-diff}. The fastest decay is observed at the alleged critical point ($D_c\simeq2$). All the curves, for varying $D$, are well-fitted by the functional form: 
\begin{equation}
    C_{\mathrm{net}}(t)\sim M(1+ct^a)^{- d},
    \label{fit-Cnet}
\end{equation}
which was empirically introduced in \cite{diffusion-birds} to measure the average fraction of non-changing neighbors after a time delay $t$ in real flocks of birds. We fitted Eq.~\eqref{fit-Cnet} on numerical data by taking as fixed parameters $d=2$ and $M=6$ (the average degree of a Voronoi vertex in a planar graph is fixed by the Euler formula). Results from the fit are shown in Fig.~\ref{fig:topo-diff}.g where parameters are plotted parametrically versus the average EPR of the system.

\subsection{Scaling with $D$}\label{sec:scaling}

Let us focus on Eq.~\eqref{resh-epr} in the two equilibrium limits $D\to0$ and $D\to\infty$: the parameter $D$ enters in the formula through $\cos(\theta_i -\theta_j )$ (alignment of bird pairs that are changing their status of neighbors), $\dot n_{ij}$ (reshuffling rate), and the $D^{-1}$ prefactor. Let us suppose that $\cos(\theta_i-\theta_j)\sim C(l)$, with $C$ the (full) spatial correlation function, which we assume isotropic, and $l$ the interaction radius (in Model I) or the average distance of a pair of birds which are leaving each other's Voronoi shell (in Model II). Let us assume that the typical reshuffling rate, $\tau_{\mathrm{resh}}\sim {\dot n_{ij}}^{-1}$, is related to the time needed for a particle to travel the same reference distance $l$:
\begin{equation}
\langle\lvert\Delta \bold x(\tau_{\mathrm{resh}})\rvert^2\rangle \sim l^2.
\label{resh-tau}
\end{equation}

In the low $D$ regime the system is deeply ordered and we can adopt the spin-wave approximation. This approximation consists in linearizing the equations of motion by taking into account only the transverse fluctuations of the velocity variables (or spin wave excitations) $v_0\boldsymbol \pi_i$, where $\boldsymbol \pi_i\perp\bold V=\frac{1}{N}\sum_{i=1}^N\bold v_i$. The magnitude of these Gaussian fluctuations scales as $|\boldsymbol\pi_i|\sim D^{1/2}$. The corresponding low-$D$ expansion of the correlation function is $C(l) \sim 1 -|\boldsymbol\pi|^2 \sim 1 - D$. In Model II $\sum_{ij} \dot n_{ij} = 0$ because the average degree of a vertex in a Delaunay triangulation of the plane (Voronoi tessellation's dual) is exactly 6 \cite{Stoyan_et_al}, so a nonzero contribution to the EPR is only expected to come from the connected correlation. In Model I a similar argument holds on average, if the system is in a stationary condition. Hence we shall replace $C(l)$ with $C_c(l)\sim D$ and estimate $\tau_{\mathrm{resh}}$ from Eq.~\eqref{resh-tau}. Since the system is strongly ordered, mutual displacement occurs mainly in the transverse space to the collective direction of motion, and we can assume that $l$ is typically small enough to work in the ballistic regime. The condition 
\begin{equation}
\langle\Delta \bold x_\perp^2(t)\rangle\sim |{\boldsymbol \pi}|^2t^2 \sim l^2
\end{equation}
identifies a reshuffling time scale $\tau_{\mathrm{resh}}\sim D^{-1/2}$, so that $\dot {\mathcal S} \sim D^{1/2}$.

In the high $D$ regime thermal noise dominates over alignment interactions. The limit model is an ideal gas of free ABPs, whose mean squared displacement is known \cite{Bar-review}:
\begin{equation}
    \langle\Delta \bold x(t)^2\rangle = \frac{2v_0^2}{D}\left[t-\frac{1}{D}\left(1-e^{-Dt}\right)\right] \sim l^2. 
    \label{MSD-ABP}
\end{equation}
From Eq.~\eqref{MSD-ABP} we deduce (both in ballistic and diffusive limits) that the reshuffling rate scales as $\tau_{\mathrm{resh}}^{-1}\sim D^{-1}$. In this disordered phase, the full correlation function corresponds to the connected one: $C_c(l)\sim e^{-l/\xi(D)}$, where $\xi(D)$ is a finite correlation length, which does not scale significantly with $D$ as $D\to\infty$. Therefore, we can assume that the mutual alignment of the considered particles will be equal to some default value, independent of $D$. The resulting scaling for the EPR is $\dot {\mathcal S} \sim D^{-2}$.

\subsection{Single particle decomposition}\label{app:A6}

\begin{figure*}[!t]
    \includegraphics[width=.45\textwidth,trim={0 0 0 1cm},clip]{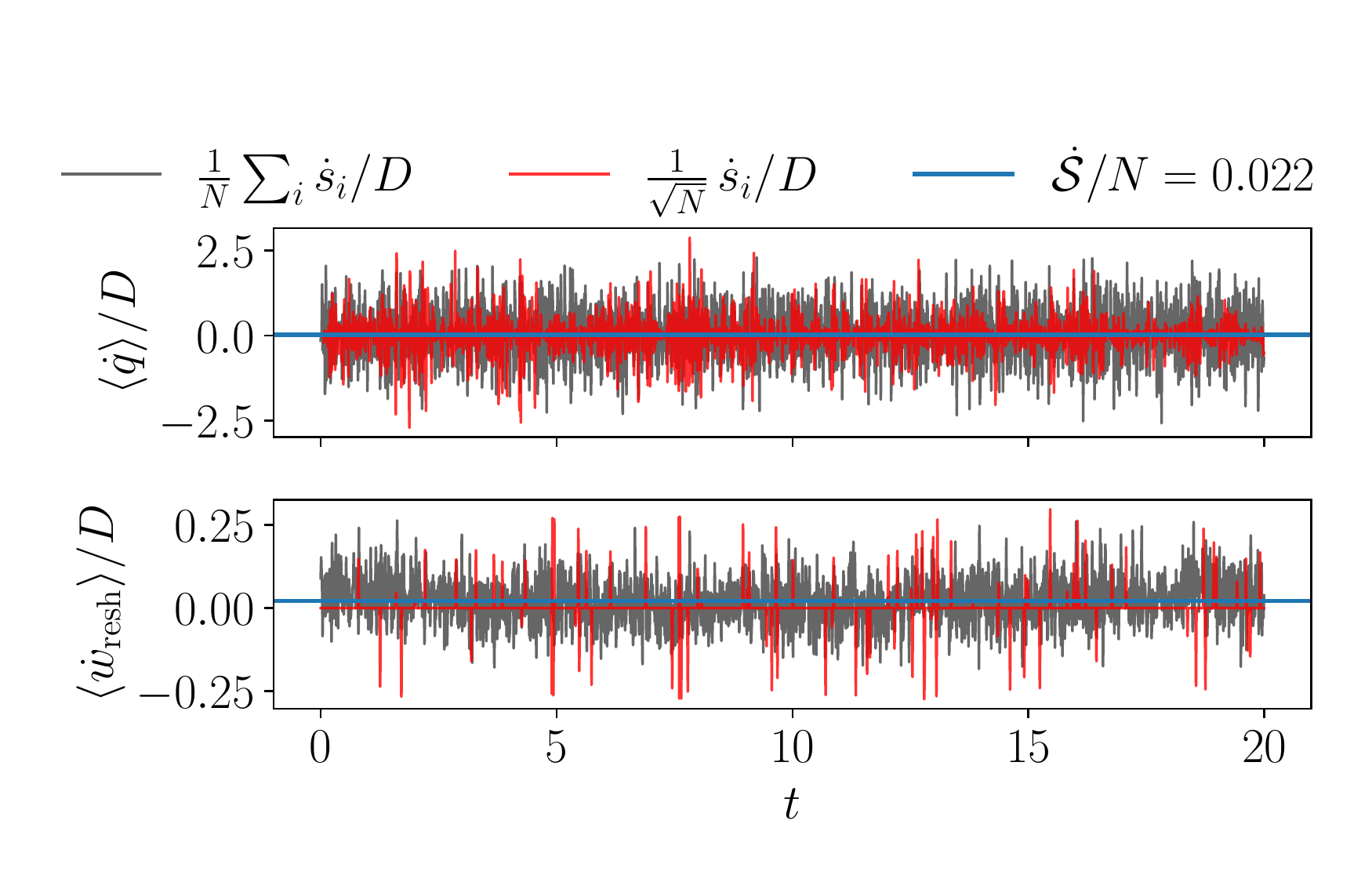}\llap{
  \parbox[b]{16cm}{\textbf{\small{a}}\\\rule{0ex}{1.72in}}}
  \hfill
    \includegraphics[width=.54\textwidth,trim={0 0 0 .5cm},clip]{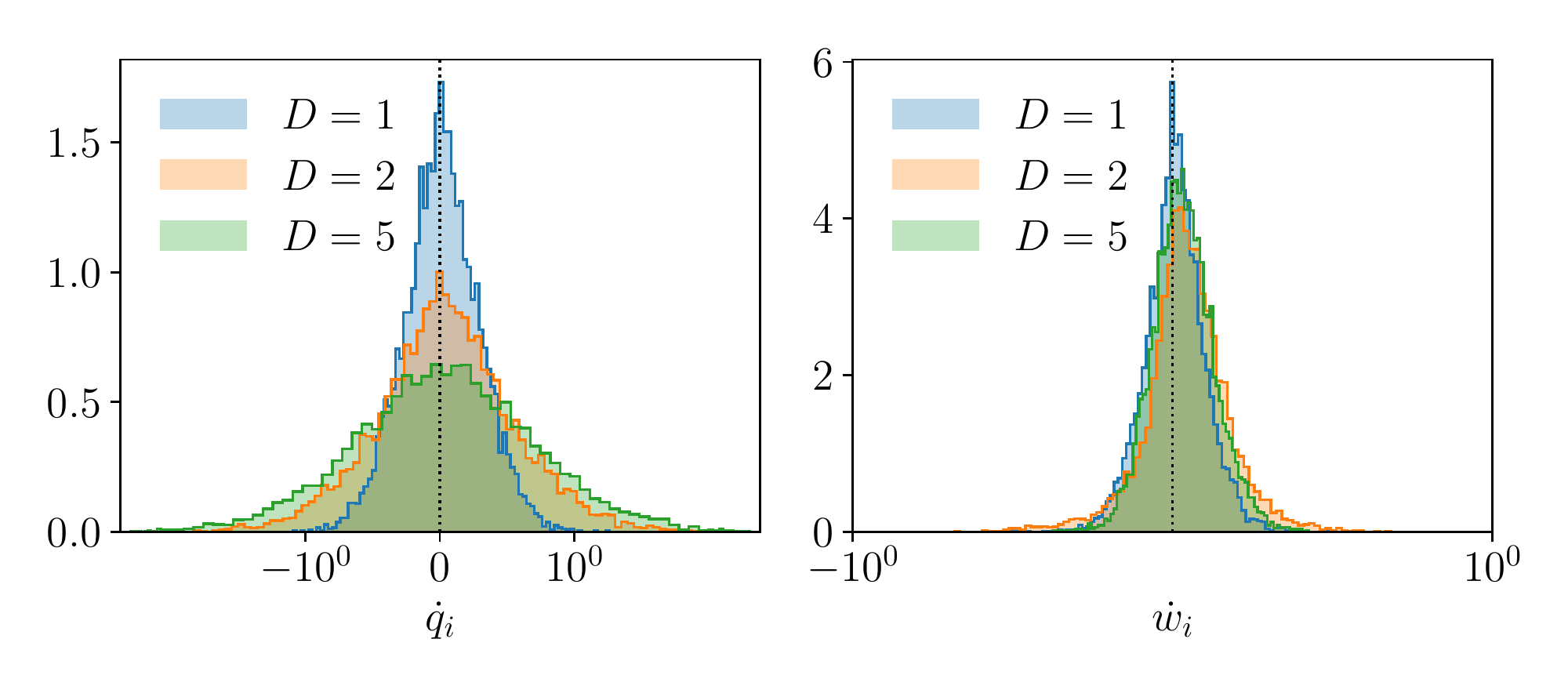}\llap{
  \parbox[b]{18.8cm}{\textbf{\small{b}}\\\rule{0ex}{1.72in}}}\llap{
  \parbox[b]{4.8cm}{\textbf{\scriptsize{$P(\dot w_i)$}}\\\rule{0ex}{1.56in}}}\llap{
  \parbox[b]{13.9cm}{\textbf{\scriptsize{$P(\dot q_i)$}}\\\rule{0ex}{1.56in}}}
    \caption{\textbf{a}: Time series of the average EPR of the flock (grey) and of the contribution of a sample single particle, properly rescaled (red), computed as in Eq.~\eqref{q-part} (top) or as in Eq.~\eqref{w-part} (bottom). The two time series have comparable fluctuations, especially when the stochastic heat is considered (top). This fact shows that all the particles contribute in the same way to the global EPR. \textbf{b}: Probability density of the single-particle stochastic heat (left, from Eq.~\eqref{q-part}) or irreversible work of fictitious reshuffling forces (right, from Eq.~\eqref{w-part}). The histogram is unimodal, even when particles are organized into polar clusters, in clear contrast with the MIPS phenomenology \cite{gonnella-work}.}
    \label{fig:non-localized}
\end{figure*}

Let us focus on Model I, where phase coexistence is realized. We consider the statistics of single-particle contributions to the EPR from Eqs.~\eqref{work-epr} and \eqref{resh-epr}, respectively corresponding to the heat rate per particle, divided by $D$:
\begin{equation}
\dot s_i = \frac{\dot q_i}{D} = -\frac{J}{D}\sum_{j}\langle\dot\theta_i\circ n_{ij}\sin(\theta_i-\theta_j)\rangle,
\label{q-part}
\end{equation}
and work of reshuffling forces per particle, divided by $D$:
\begin{equation}
\dot s_i = \frac{\dot w_i}{D} = -\frac{J}{2D}\sum_{j}\langle\dot n_{ij}\circ \cos(\theta_i-\theta_j)\rangle.
\label{w-part}
\end{equation}
The goal of this analysis is to unveil possible correlations between dissipation and spatial segregation of each self-propelled particle, as it has been observed in scalar active matter.

In contrast to MIPS models \cite{PRX-mips-entropy,revAOUP,gonnella-work}, however, dissipation seems to occur in a non-localized way in the system.
Two findings have been made in the context of scalar active matter:
first of all, the analysis of both coarse-grained models and microscopic agent-based models for MIPS has shown that the breakdown of the time-reversal symmetry occurs at the interfaces, whereas in the bulk of both the dense and dilute phases the EPR density vanishes \cite{revAOUP,PRX-mips-entropy}. A second analysis on a different type of microscopic model (ABP with repulsive interactions, in contrast to Active Ornstein-Uhlenbeck particles) has revealed that the work of the self-propulsion force per single particle assumes a different value depending on the particle's position inside the system \cite{gonnella-work}. The probability distribution of the active work per particle is assumed to obey a large deviation principle and the associated rate function is found to be non-convex. The values corresponding to the minima of the rate function are the typical active work values of particles belonging to the bulk of the two phases; work values in the non-convex region are typical of particles at the interface.

If we represent the EPR contribution per particle, using either Eq.~\eqref{q-part} or Eq.~\eqref{w-part}, we see that violations of the time reversal symmetry are not concentrated at the interface; on the contrary, the bulk contributes significantly (not with a definite sign) --- see supplementary movies. If we also consider the time series of single-particle EPR contributions, we see that they largely fluctuate, taking both positive and negative values. The typical amplitude of fluctuations is comparable to that of the rescaled average EPR of the system, $\frac{1}{\sqrt{N}}\sum_{i=1}^Ns_i$ (Fig.~\ref{fig:non-localized}.a). Finally, the histograms of the time-averaged single particle contributions, shown in Fig.~\ref{fig:non-localized}.b, seem to have a unimodal shape, corroborating the idea that in polar flocks there is no correspondance between the particle's energetic role and its positioning in the flock.

\section{}\label{app:B}

In this section we want to analyze general consequences deriving from the irreversibility condition on ABP models with alignment.
We start by recalling some results presented in \cite{DalCengio} and move from them to:
\begin{enumerate}
\item derive a new formula for the entropy production of non-equilibrium stochastic process driven by additive noise (Eq.~\eqref{epr-new});
\item show that the irreversibility condition induces on any ABP-based model of flocking an explicit asymmetry in the steady state distribution of the system's microstates; 
\item generalize this last result to any Langevin system with completely irreversible drift.
\end{enumerate}

\subsection{General results}
Let us consider a general stochastic additive process $\bold z(t)$ described by a Langevin equation with drift term $\bold A(\bold x)$ and diffusion term $\bold D$. Let $\psi(\bold z)$ be the steady-state distribution that solves the stationary Fokker-Planck equation: $\mathcal L_{FP}\psi(\bold z) = - \nabla \cdot\bold j(\bold z) = 0$, where 
\begin{equation}
\bold j(\bold z) = \bold A(\bold z)\psi(\bold z) - \bold D\nabla \psi(\bold z) 
\label{current}
\end{equation}
is the probability current. This current is standardly decomposed into a \emph{reversible} and \emph{irreversible} part \cite{Spinney-Ford} as:
\begin{flalign}
    \bold j^{rev} (\bold z) &= \bold A^{rev}(\bold z)\psi(\bold z) ,\\  
    \bold j^{irr}(\bold z) &= \bold A^{irr}(\bold z)\psi(\bold z) - \bold D\nabla \psi(\bold z),
    \label{def-j-rev}
\end{flalign}
where: 
\begin{flalign}
	A_{\alpha}^{rev}(\bold z) &= \frac{1}{2}\left[A_{\alpha}(\bold z) - \epsilon_{\alpha}A_{\alpha}(\boldsymbol\epsilon\bold z)\right], \\
	A_{\alpha}^{irr}(\bold z) &= \frac{1}{2}\left[A_{\alpha}(\bold z) + \epsilon_{\alpha}A_{\alpha}(\boldsymbol\epsilon\bold z)\right],
\end{flalign}
with $\boldsymbol\epsilon$ denoting the time reversal operator. For now, we assume that $\boldsymbol\epsilon$ acts linearly on the state variable $\bold z$ (precisely, diagonally with $\epsilon_{\alpha} = \pm1$ $\forall \alpha$).

A necessary and sufficient condition for detailed balance to hold is that the two following conditions are verified (in addition to stationarity, $\nabla \cdot \bold j=0$) \cite{Gardiner}:
\begin{equation}
    \bold j^{irr}(\bold z)=0 \quad \text{and} \quad D_{\alpha\beta}(\bold z) = \epsilon_{\alpha}  \epsilon_{\beta}D_{\alpha\beta}(\boldsymbol\epsilon\bold z).
    \label{DB}
\end{equation}
Therefore the condition of irreversibility implies $\bold j^{irr}(\bold z) \neq0$, when $\bold D$ is independent of $\bold z$. This condition is sufficient to guarantee the positivity of the entropy production in a proper NESS, i.e. in the absence of external drivings, where the total entropy production reduces to the housekeeping entropy production:
\begin{equation}
    \dot{\mathcal S}^{hk} = \int d\bold z\, \psi(\bold z)\, V^{irr}_{\alpha}(\boldsymbol\epsilon\bold z)\left(\bold D^{-1}(\boldsymbol\epsilon\bold z)\right)_{\alpha\beta} V^{irr}_{\beta}(\boldsymbol\epsilon\bold z).
    \label{epr-hk}
\end{equation}
In \eqref{epr-hk} $\bold V(\bold z)=\bold j(\bold z)/\psi(\bold z)$ is the phase space velocity \cite{Spinney-Ford}, which is decomposed as $\bold V^{rev}(\bold z)+\bold V^{irr}(\bold z)$ following \eqref{def-j-rev}. In \cite{DalCengio} Dal Cengio et al. identified Eq.~\eqref{epr-hk} with a second expression for the EPR given in \cite{Spinney-Ford},
\begin{equation}
    \dot {\mathcal S} = \int d\bold z\, \psi(\bold z)\, V^{irr}_{\alpha}(\bold z)\left(\bold D^{-1}(\bold z)\right)_{\alpha\beta} V^{irr}_{\beta}(\bold z),
    \label{epr-SF}
\end{equation} 
to derive the following constraint:
\begin{equation}
    \sum_{\alpha\beta}{D^{-1}}_{\alpha\beta}V_{\alpha}^{irr}(\bold z){V_{\beta}^{irr}(\bold z)} = \sum_{\alpha\beta}{D^{-1}}_{\alpha\beta} V_{\alpha}^{irr}(\boldsymbol\epsilon\bold z){V_{\beta}^{irr}(\boldsymbol\epsilon\bold z)}. 
    \label{constraint-V}
\end{equation}
Eq.~\eqref{constraint-V} is valid almost surely if the diffusion matrix is invertible and $\bold z$-independent. Using the definition of $\bold V^{irr}$ and the property $A_{\alpha}^{irr}(\bold z) = \epsilon_{\alpha}A_{\alpha}^{irr}(\boldsymbol \epsilon \bold z)$, Eq.~\eqref{constraint-V} implies: 
\begin{equation}
    \sum_{\alpha} \left[A_{\alpha}^{irr}(\bold z) + D_{\alpha\beta}\partial_{\beta}\phi_{+}(\bold z)\right]\partial_{\alpha}\phi_{-}(\bold z) = 0
    \label{SaraDC}
\end{equation}
almost surely. The functions $\phi_{\pm}$ are defined as the T-symmetric and T-antisymmetic parts of the quasi-potential $\phi(\bold z) = -\log \psi(\bold z)$:
\begin{equation}
    \phi_{+}(\bold z) = \frac{1}{2}\left[\phi(\bold z)+\phi(\boldsymbol\epsilon\bold z)\right], \qquad \phi_{-}(\bold z)  = \frac{1}{2}\left[\phi(\bold z)-\phi(\boldsymbol\epsilon\bold z)\right].
    \label{def-sym-asm-qp}
\end{equation}

For the sake of simplicity, let us assume $\bold D$ is diagonal. All the results can be generalized to the non-diagonal case, if $\bold D$ is symmetric and positive definite. We remark that it is not possible to merely invoke the diagonalizability of $\bold D$ through a change of basis, because we need all the coordinates of $\bold z$ to have a definite parity under time reversal. 

We start from an explicit rewriting of the entropy production rate from Eq.~\eqref{epr-SF}:
\begin{equation}
    \dot{\mathcal S} = \int d\bold z\,  e^{-\phi(\bold z)} \sum_{\alpha} D^{-1}_{\alpha\alpha} \left[A_{\alpha}^{irr}(\bold z) + D_{\alpha}\partial_{\alpha}\phi(\bold z)\right]^2.
    \label{start-epr-gen}
\end{equation}
Exploiting the symmetries of $\bold A^{irr}$, $\phi_{+}$ and $\phi_{-}$, we can rewrite Eq.~\eqref{start-epr-gen} as
\begin{widetext}
\begin{dmath}
    \dot{\mathcal S} = \int d\bold z\, 2\cosh\left(\phi_{-}(\bold z)\right) e^{-\phi_{+}(\bold z)} \sum_{\alpha} D^{-1}_{\alpha\alpha} \left[\left(A_{\alpha}^{irr}(\bold z) + D_{\alpha\alpha}\partial_{\alpha}\phi_{+}(\bold z)\right)^2 + D_{\alpha\alpha} \left(\partial_{\alpha}\phi_-(\bold z)\right)^2\right] + \int d\bold z\, 4\sinh\left(\phi_{-}(\bold z)\right) e^{-\phi_{+}(\bold z)} \sum_{\alpha}\left(A_{\alpha}^{irr}(\bold z) + D_{\alpha\alpha}\partial_{\alpha}\phi_{+}(\bold z)\right)\partial_{\alpha}\phi_-(\bold z).
    \label{medium-epr-gen}
\end{dmath}
The second line of Eq.~\eqref{medium-epr-gen} is zero thanks to Eq.~\eqref{SaraDC}, hence: 
\begin{equation}
    \dot{\mathcal S} = \int d\bold z\, \cosh\left(\phi_{-}(\bold z)\right) e^{-\phi_{+}(\bold z)} \sum_{\alpha} \left[D^{-1}_{\alpha\alpha} \left(A_{\alpha}^{irr}(\bold z) + D_{\alpha\alpha}\partial_{\alpha}\phi_{+}(\bold z)\right)^2 + \left(\partial_{\alpha}\phi_-(\bold z)\right)^2\right]\geq0.
    \label{epr-new}
\end{equation}
\end{widetext} 
Equality is realized in Eq.~\eqref{epr-new} iff both of the following conditions are verified: 
\begin{equation}
    A_{\alpha}^{irr}(\bold z) + D_{\alpha}\partial_{\alpha\alpha}\phi_{+}(\bold z) =0 \quad \forall \alpha
    \label{cond1B}
\end{equation} 
and 
\begin{equation}
    \partial_{\alpha}\phi_-(\bold z) = 0\quad \forall \alpha \iff \phi_-(\bold z) =0.
    \label{cond2B}
\end{equation}
Notice that $\phi_{-}$ cannot be a constant function different from zero.

Breakdown of detailed balance imposes that at least one of the terms in Eqs.~\eqref{cond1B}--\eqref{cond2B} is nonzero, along with the constraint \eqref{SaraDC}. If all the coordinates of the state variable are T-even, then $\phi_-(\bold z)=0$ and violation of Eq.~\eqref{cond1B} is enforced. In the presence of T-odd coordinates, $\phi_-(\bold z)$ can be different from zero. 

\subsection{Application to Langevin-Vicsek model}
The results presented above are valid for general Langevin stochastic processes, provided that they are additive with an invertible diffusion matrix.
Let us now focus on our system of ABPs in Eqs.~\eqref{eq-x}--\eqref{eq-th}, which models a Langevin-Vicsek flock. 

Since we work in the absence of translational diffusion, the $\bold D$ matrix is non-invertible, irrespectively of whether we decide to study the process in the $(\bold X, \boldsymbol\Theta)$ phase space or in the $(\bold X, \bold V)$ phase space, where $\bold V=e^{i\boldsymbol\Theta}$. Moreover, the process is non-additive in the second case. In the first case the $\boldsymbol\Theta$ variables are not T-even nor T-odd under time reversal, but the time reversal operator acts by shifting them of an amount $\pi$. Hence the hypotheses that led to the derivation of the above results are not valid for the polar system of interest. Nonetheless, we can still derive from the irreversibility condition a condition on the asymmetry of the steady-state distribution $\psi(\bold z)$. 

Let us apply the same definition of quasi-potential $\phi(\bold z)$ and of its decomposition into T-symmetric and T-antisymmetric parts given in \eqref{def-sym-asm-qp}. Our goal is to show (by contradiction) that irreversibility implies $\phi_-(\bold z)\neq0$. 
Let us start from the stationary FPE,
\begin{widetext}
\begin{equation}
    v_0 \sum_i\bold e(\theta_i)\cdot \nabla_i \psi = J\sum_{ij}n_{ij}\partial_{\theta_i}(\sin(\theta_i-\theta_j)\psi) + D \sum_i\partial^2_{\theta_i\theta_i}\psi,
	 \label{FPE-vicsek}
\end{equation}
and derive the corresponding PDE for the quasi-potential $\phi$:
\begin{dmath}
    v_0 \sum_i\bold e(\theta_i)\cdot \nabla_i \phi = J\sum_{ij}n_{ij}\cos(\theta_i-\theta_j) + J\sum_{ij}n_{ij}\sin(\theta_i-\theta_j)\partial_{\theta_i}\phi + D \sum_i\partial^2_{\theta_i\theta_i}\phi + D \sum_i(\partial_{\theta_i}\phi)^2.
    \label{FPE-phi}
\end{dmath}
Let us apply the time-reversal operator to both the R.H.S. and L.H.S. of Eq.~\eqref{FPE-phi} (this operation corresponds to a change of variable from the state variable $\bold z$ to $\boldsymbol\epsilon\bold z$) and split the quasi-potential into its T-symmetric and T-antisymmetric parts to rewrite a set of two coupled stationary equations:
\begin{dmath}
    v_0 \sum_i\bold e(\theta_i)\cdot \nabla_i \phi_- = J\sum_{ij}n_{ij}\cos(\theta_i-\theta_j) + J\sum_{ij}n_{ij}\sin(\theta_i-\theta_j)\partial_{\theta_i}\phi_+ + D \sum_i\partial^2_{\theta_i\theta_i}\phi_+ + D\sum_i(\partial_{\theta_i}\phi_+)^2 + D\sum_i(\partial_{\theta_i}\phi_-)^2;
    \label{FPphi+}
\end{dmath}
\begin{equation}
    v_0 \sum_i\bold e(\theta_i)\cdot \nabla_i \phi_+ = J\sum_{ij}n_{ij}\sin(\theta_i-\theta_j)\partial_{\theta_i}\phi_- + D \sum_i\partial^2_{\theta_i\theta_i}\phi_- + 2D \sum_i(\partial_{\theta_i}\phi_+)(\partial_{\theta_i}\phi_-).
    \label{FPphi-}
\end{equation}
\end{widetext}
We now assume that $\phi_-(\bold z)=0$ in Eq.\eqref{FPphi+}. The resulting equation for $\psi = e^{\phi_+}$ is of the form:
\begin{equation}
    0=J\sum_{ij}n_{ij}\partial_{\theta_i}(\sin(\theta_i-\theta_j)\psi) + D \sum_i\partial^2_{\theta_i\theta_i}\psi.
    \label{eq-FPE}
\end{equation}
This equation is solved by the Boltzmann distribution that we would have in the absence of self-propulsion:
\begin{equation}
    \psi(\bold z)=f(\bold X)\exp[-\mathcal H(\bold z)] \iff \phi_+(\bold z) = \frac{\mathcal H(\bold z)}{D} + c(\bold X),
    \label{Boltzmann-eq}
\end{equation}
with $c(\bold X)$ an arbitrary constant. Uniqueness of the stationary solution of a Fokker-Planck equation is guaranteed under rather general smoothness hypotheses on the solution and drift term \cite{Zeeman}. Let us notice that we crucially exploited the symmetry of $n_{ij}$ to write the steady state solution as Eq.~\eqref{Boltzmann-eq}. We can conclude that the non-equilibrium condition is contradicted: the hypothesis $\phi_-(\bold z)=0$ leads to an absurdum.  

In Langevin-Vicsek models, the condition $\phi_-(\bold z)\neq0$ corresponds to requiring that 
\begin{equation}
\psi(\bold X, \bold V)\neq \psi(\bold X, e^{i\pi}\bold V),
\label{asymm}
\end{equation}
i.e. irreversibility constrains an explicit symmetry breaking under rotations in the internal space of velocities. A possible way to measure of the degree of irreversibility of the process is to quantify the asymmetry of the steady state pdf in Eq.~\eqref{asymm}. However, reconstructing the $N$-body probability density is clearly an out-of-reach task, both in numerical simulations or real experiments. Having some knowledge of the aligning interaction potential $\mathcal H_{XY}(\boldsymbol\Theta;\bold n(X))$ allows us to visualize the asymmetry on a much lower-dimensional space, and to predict how the second law of thermodynamics ($\dot {\mathcal S} \geq0$) constrains the realization of such asymmetries.

\subsection{Langevin processes with irreversible drift}
The result obtained in the previous section can be extended to any Langevin process where $\bold A^{rev}(\bold z)=0$ and the state variable coordinates have a definite parity under time reversal. It is sufficient to write the stationary Fokker-Planck equation
\begin{equation}
\nabla\cdot\bold A + \bold A\cdot \nabla \phi + D \nabla^2\phi + D\left(\nabla\phi\right)^2 =0
\label{FPE-gen}
\end{equation}
and perform a change of variable, identifying $\bold z$ as $\boldsymbol\epsilon\bold z'$. Using the decomposition of $\bold A$ and $\phi$ into their T-symmetric and T-antisymmetric components, and combining (respectively with a positive and negative sign) the resulting equation with \eqref{FPE-gen}, one can obtain the two following equations:
\begin{widetext}
\begin{flalign} 
\label{abs1} \nabla_O\cdot \bold A_O^{irr} + \bold A_E^{irr}\cdot\nabla_E\phi_- + \bold A_O^{irr}\cdot\nabla_O\phi_+ +  D\nabla^2\phi_+ + D\left(\nabla\phi_+\right)^2 + D\left(\nabla\phi_-\right)^2&=0\\
\label{abs2}\nabla_E\cdot \bold A_E^{irr} + \bold A_E^{irr}\cdot\nabla_E\phi_+ + \bold A_O^{irr}\cdot\nabla_O\phi_- + D\nabla^2\phi_- + 2D\nabla\phi_+\cdot\nabla\phi_-&=0.
\end{flalign} 
\end{widetext}
The subscripts in the equations above indicate the parity of the state variable coordinates, which can be split as $\bold z = (\bold z_E,\bold z_O)$, such that $\boldsymbol\epsilon \bold z = (\bold z_E, -\bold z_O)$. Assuming that $\phi_-(\bold z)=0$ (absurdum), the sum of \eqref{abs1} and \eqref{abs2} yields
\begin{equation}
\nabla \cdot \left(\bold A^{irr}\psi\right) + D\nabla^2\psi=0,
\label{abs-gen}
\end{equation} 
where we have identified $\psi=e^{\phi_+}$. Providing natural boundary conditions to the FPE \eqref{FPE-gen}, we deduce from \eqref{abs-gen} that $\bold A^{irr}\psi + D\nabla\psi=0$. This condition, together with $\phi_-(\bold z)=0$, implies detailed balance. Therefore irreversibility implies, in a system where $\bold A^{rev}=0$ and the state variables coordinates are either T-odd or T-even, that $\phi_-(\bold z)\neq0$.

\bibliography{bibliography}

\end{document}